\documentclass{article}
\usepackage[english]{babel}
\usepackage[utf8]{inputenc}
\usepackage{johd}
\usepackage{amsmath}
\usepackage{algorithm}
\usepackage{algpseudocode}
\usepackage{graphicx}
\usepackage{johd}
\usepackage{amssymb}
\usepackage{amsfonts}
\usepackage{verbatim}
\usepackage{lscape}
\usepackage{multirow}

\DeclareMathOperator*{\argmin}{arg\,min}

\usepackage{calrsfs}
\DeclareMathAlphabet{\pazocal}{OMS}{zplm}{m}{n}

\title{Spectral identification and estimation of mixed causal-noncausal invertible-noninvertible models}

\author{Alain Hecq\footnote{a.hecq@maastrichtuniversity.nl, Maastricht University, SBE, Department of Quantitative Economics, P.O.Box 616, 6200 MD Maastricht, The Netherlands.} $\;$ and Daniel Velasquez-Gaviria\footnote{Corresponding author: d.velasquezgaviria@maastrichtuniversity.nl, Maastricht University, SBE, Department of Quantitative Economics, P.O.Box 616, 6200 MD Maastricht, The Netherlands.}\\
        \small Maastricht University\\
}

\date{\today} 

\begin{document}
\linespread{1.5}
\maketitle

\begin{abstract}
    This paper introduces new techniques for estimating, identifying and simulating mixed causal-noncausal invertible-noninvertible models. We propose a framework that integrates high-order cumulants, merging both the spectrum and bispectrum into a single estimation function. The model that most adequately represents the data under the assumption that the error term is i.i.d. is selected. Our Monte Carlo study reveals unbiased parameter estimates and a high frequency with which correct models are identified. We illustrate our strategy through an empirical analysis of returns from 24 Fama-French emerging market stock portfolios. The findings suggest that each portfolio displays noncausal dynamics, producing white noise residuals devoid of conditional heteroscedastic effects. 
\end{abstract}

\noindent\\



\section{Introduction}

Econometricians have traditionally considered ARMA($p,q$) models using the Box-Jenkins methodology to describe and forecast time series. This method heavily rely on the assumption that the autoregressive (AR) and moving average (MA) components should be respectively causal and invertible. This essentially means that the roots of both AR and MA polynomials are situated outside the unit circle on the complex plane, which guarantees that the model converges to a unique, stable solution. Instead, non-causal and non-invertible models have their respective roots located within the unit circle. The increasing interest in these models comes from their adeptness at capturing complex non-linear phenomena such as bubbles and asymmetric cycles. This capacity to reflect anticipation and speculation resonates strongly with observable behaviors in financial markets (\cite{lanne2011noncausal}, \cite{lof2013noncausality,hecq2016identification,lof2017noncausality,hecq2022spectral, hecq2019predicting}, \cite{gourieroux2017local}). 

The interpretation of noncausal and noninvertible models owns a lot to the literature looking at non-fundamental solutions within expectation frameworks, suggesting that expectations can significantly influence present-day prices. In \cite{muth1961rational,lucas1972expectations,fama1970efficient,fischer1977long}, within the paradigm of the rational expectations hypothesis, agents judiciously harness all accessible information when making consumption and investment decisions. This includes forecasts about prospective events. A comprehensive understanding of how future information affects current prices emerges from \cite{hansen1991two}. The main argument is the disparity between the breadth of information available to agents versus econometricians, advocating for the incorporation of future values, or at the very least, expectations of such values, to bolster the precision of models. In \cite{diba1988theory}'s "rational bubble" framework stock prices are not solely anchored to conventional market indicators such as anticipated dividends but also by elements not rooted in market fundamentals. In feedback loop models (see \cite{shiller1984stock}) traders that are foreseeing an uptick in a stock's value, may engage in contemporary buying sprees. 

To emulate such intricate behaviors, we identify and estimate mixed autoregressive causal-noncausal invertible-noninvertible moving average model, denoted as MARMA($r,s,r',s'$). This model decomposes the AR($p$) and MA($q$) polynomials based on the positioning of their roots relative to the unit circle. The outcome is a refined ARMA model, which considers both lags and leads of the endogenous variable, where the relationship $p=r+s$. Furthermore, this model incorporates both past (retrospective) and future (anticipatory) errors, represented by the equation $q=r'+s'$. MARMA models have a unique and stationary solution (\cite{brockwell2009time} (p. 81)). Notably, this deviates from the nonstationary solutions where the endogenous variable remains contemporaneously uncorrelated with its error sequence. Traditional estimation techniques, grounded in Gaussianity or second-order moments, fail to pinpoint this stationary solution. This stems from the fact that both causal (invertible) and noncausal (noninvertible) processes exhibit identical autocorrelation functions and spectral densities. Therefore, the resultant estimated coefficients predominantly have roots outside the unit circle, disregarding the possibility that the true process structure might possess roots within the unit circle. Indeed, an ARMA($p,q$) model has $2^{p+q}$ parameter configurations with identical probability structures (\cite{rosenblatt2012random}). However, when the error term follows a non-Gaussian distribution, each of the $2^{p+q}$ process is distinctly identifiable within a MARMA($r,s,r',s'$) representation (\cite{wu2010least}). The repercussions for an econometrician's misstep in estimating a time series that potentially integrates noncausal or noninvertible dynamics using traditional ARMA models can have important modelling consequences. As such one can get spurious higher-order dependencies in the conditional moments of errors, which is akin to conditional heteroscedasticity (manifesting as GARCH-type effects; see \cite{gourieroux2013explosive}, proposition 3), \cite{rosenblatt2000gaussian, velasco2022estimation}).

 Most literature on MAR (and MARMA) gravitates towards the Maximum Likelihood estimator when exploring error sequences that are both i.i.d and non-Gaussian (\cite{lii1982deconvolution,lii1992approximate,lii1996maximum, lanne2011noncausal, hecq2016identification, meitz2013maximum, breid1991maximum, huang2000quasi}). MLE boasts consistency across symmetric, strictly positive, and smooth non-Gaussian probability density functions. Yet, empirical situations often lack prior knowledge of distribution functions. Any arbitrary choice in this context can potentially introduce biases in parameters and erroneous model identification. Our paper offers an innovative approach by sidestepping this requirement. Instead, we advocate a spectral domain strategy using higher-order cumulants and spectral densities within a least squares framework. Our method hinges on the minimum distance estimation between the second and third-order spectral densities (namely, the spectrum and bispectrum). Using the frequency domain representation of the data via the periodogram and biperiodogram, we minimize the second and third-order dependence, aligning with the white noise and i.i.d premises, moreover  obtaining error sequences that do not exhibit heteroscedastic effects. Note that \cite{rosenblatt1980linear} already recommended an amalgamation of both the spectrum and bispectrum when considering ARMA models with roots outside or within the unit circle. See also \cite{lii1982deconvolution}, \cite{kumon1992identification} and \cite{brillinger1985fourier} for higher-order spectral methodologies. Closer to our paper, \cite{velasco2018frequency} incorporate spectral densities and periodograms exceeding the third order into the estimation matrix of potential noncausal and noninvertible ARMA processes. See also \cite{lobato2022single} for non-stationary fractional integrated ARMA process models. 

Our paper presents several contributions to the study of MARMA models. We first introduce a variation of the estimation method proposed in \cite{velasco2018frequency}. We modify the transfer function to distinguish between the polynomials with roots inside and outside the unit circle, tailored for MARMA models. This draws a clear distinction between causal-noncausal and invertible-noninvertible polynomials, a strategy reminiscent of \cite{lii1996maximum}. \cite{hecq2022spectral} observed a pronounced reliance on initial values. To remedy this, our current approach directly targets parameter estimation using Genetic Algorithms rather than their root-based factorization, significantly reducing dependence on initial values. The advantage of using Genetic Algorithms is that by simulating the evolutionary process, they explore a vast parameter space, rendering initial values irrelevant and enhancing the likelihood of finding the global minimum of the target function.

Notably, our refined estimator's consistency and asymptotic normality mirror those of its predecessor. Secondly, we develop an identification of the MARMA model based on the existence of the global minimum of our estimation function. This minimum remains consistent, provided the second and third-order cumulants are present. Our third contribution is an alternative method to simulate MARMA processes in the frequency domain. We exploit the inverse Fourier transform of the error sequence combined with the model's transfer function, ensuring streamlined replication. Our fourth contribution is the study of the monthly returns across 24 Fama-French portfolios in emerging markets. Drawing a parallel to \cite{gospodinov2015minimum}, which unearthed a noninvertible MA(0,1) model across 25 similarly curated portfolios in developed markets, our findings reveal a mosaic of noncausal and mixed causal-noncausal dynamics. To test the validity of the i.i.d property on our derived residuals, we deploy statistical tools proposed by \cite{dalla2020robust}. We consistently obtain error sequences devoid of heteroscedastic effects under our identified and estimated noncausal dynamics.

The paper is divided as follows: Section 2 presents our methodology. Section 3 delves deep into spectral estimation. Section 4 introduces i.i.d tests. Section 5 delineates our MARMA model simulation procedure. Sections 6 and 7, respectively, showcase the insights from our Monte Carlo analysis and the empirical application. The final section concludes.

\section{Methodology}

\subsection{ARMA($p,q$) model}

The zero mean (for the simplicity of the presentation) stationary time series $y_t$ for every $t=\left\{0,\pm 1, \pm2, \pm 3, ... \right\}$ with length $T$ and $\mathbb{E}\left[y_t^2\right]<\infty$, is said to be an ARMA($p,q$) process if it satisfies the recursion

\begin{equation}\label{arma1}
    y_t-\phi_1y_{t-1}-...-\phi_py_{t-p}=\epsilon_t+\theta_{1}\epsilon_{t-1}+...+\theta_q\epsilon_{t-q},
\end{equation}
where $\epsilon_t\sim \textrm{WN}(0,\sigma^2)$ is white noise. Note that this definition only considers the first two moments of $\epsilon_t$. Equation \ref{arma1} can be written in a compact form as 

\begin{equation}
    \phi(L)y_t=\theta(L)\epsilon_t,
\end{equation}
where $L$ is the backward shift operator, that is, $L^\ell=y_{t-\ell}$ for $\ell=\left\{0,\pm 1, \pm2, \pm 3, ... \right\}$, $\phi(L)$ is a $p$-th order autoregressive (AR) polynomial, and $\theta(L)$ is a $q$-th order moving-average (MA) polynomial. Both polynomials can be expressed in terms of their roots as

\begin{equation}
    \phi(z)=1-\phi_1z-...-\phi_pz^{p},\;\; \textrm{and}\;\; \theta(z)=1+\theta_1z+...+\theta_qz^{q},
\end{equation}
where $\phi(z)$ and $\theta(z)$ have no common roots, and all of their roots are outside the unit circle, that is $\phi(z)\neq0\;\;\textrm{for}\;\; \lvert z\rvert<1$, and $\theta(z)\neq0\;\;\textrm{for}\;\; \lvert z\rvert<1$. Therefore, there exists a unique and strictly stationary solution of ARMA($p,q$), determining that the model is causal and invertible. The ARMA($p,q$) model is causal if there exists a sequence of constants $\left\{\pi_j\right\}$, such that

\begin{equation}
    y_t=\sum_{j=0}^{\infty}\pi_j\epsilon_{t-j},
\end{equation}
with $\sum^{\infty}_{0}\lvert \pi_j \rvert<\infty$ and $y_t$ expressed in terms of the present and past of the errors sequence that is an MA($\infty$). On the other hand, the ARMA($p,q$) model is invertible if there exists a sequence of constants $\left\{\pi^{*}_j\right\}$, such that

\begin{equation}
    \epsilon_t=\sum_{j=0}^{\infty}\pi^{*}_jy_{t-j},
\end{equation}
with when \(\sum_{j=0}^{\infty} \lvert \pi^{*}_j \rvert < \infty\), \(\epsilon_t\) is formulated in terms of both the present and historical values of \(y_t\), effectively characterizing it as an AR(\(\infty\)) process. The nature of the ARMA(\(p,q\)) model is determined by the roots of the polynomials \(\phi(z)\) and \(\theta(z)\). Specifically, if the roots of \(\phi(z)\) lie within the unit circle, the model is categorized as noncausal. Conversely, the presence of the roots of \(\theta(z)\) within the unit circle designates the model as noninvertible.

For time series where \(y_t\) adheres to Gaussian properties, its entire probability framework is anchored to its autocovariance function, see \cite{lii1996maximum}. A consequence of this property is the existence of \(2^{p+q}\) distinct models, each formulated from different parameter combinations, yet sharing an identical autocovariance function. This overlap obfuscates the differentiation between noncausal and noninvertible models; second-order estimators invariably converge to their causal and invertible analogs. As a result, it has become a prevailing practice among econometricians to sidestep noncausal and noninvertible models, especially given that no substantial insights are forsaken from a second-order viewpoint, as discussed in \cite{brockwell2009time}, p. 81. However, in the context of non-Gaussian \(y_t\), each model manifests with a distinct probability structure, facilitating the discernment and identification of noncausal and noninvertible models when they emerge. Such precision invariably leads to an enhanced data representation, as elaborated by \cite{wu2010least}.

\subsection{MARMA($r,s,r',s'$) model}

The ARMA($p,q$) model can be extended to obtain the mixed causal-noncausal and invertible-noninvertible process, namely the MARMA($r,s,r',s'$) model

\begin{equation}\label{arma2}
    (1-\phi_1^{+}L-...-\phi_r^{+}L^r)(1-\phi_1^{*}L^{-1}-...-\phi_s^{*}L^{-s})y_t=(1+\theta^{+}_1L+...+\theta_{r'}^{+}L^{r'})(1+\theta^{*}_1L^{-1}+...+\theta^{*}_{s'}L^{-s'})\varepsilon_t,
\end{equation}
where $\varepsilon_t \overset{\mathrm{i.i.d}}{\sim} \mathfrak{G}(\zeta)$ with $\mathfrak{G}$ a distribution function and $\zeta$ its parameters. The first three moments/cumulants are denoted

\begin{equation}
    \mathbb{E}(\varepsilon_t)=0,\;\;\mathbb{E}(\varepsilon_t^2)=\kappa^{*}_2,\;\;\mathbb{E}(\varepsilon_t^3)=\kappa^{*}_3.
\end{equation}

Moreover, $\varepsilon_t$ has bounded moments of order $k$, with $k\geq 3$. Note that in the traditional ARMA model, $\epsilon_t$ is white noise, implying that all higher-order properties of its joint distribution are ignored, except for its first and second moments, $E(\epsilon_t)$ and $E(\epsilon_t\epsilon_s)$. On the other hand, $\varepsilon_t$ is i.i.d, implying independence in its high-order moments. Equation \ref{arma2} can be expressed in terms of its decomposed polynomials as 

\begin{equation} \label{dec}
\begin{split}
\phi(z) & = \phi^{+}(z)\phi^{*}(z)=(1-\phi^{+}_1z-...-\phi^{+}_rz^r)(1-\phi^{*}_1z-...-\phi^{*}_sz^s), \\
 \theta(z) & = \theta^{+}(z)\theta^{*}(z)=(1+\theta^{+}_1z+...+\theta^{*}_{r'}z^{r'})(1+\theta^{*}_1z-...-\theta^{*}_{s'}z^{s'}).
\end{split}
\end{equation}

The polynomials $\phi^{+}(z)$, $\phi^{*}(z)$, $\theta^{+}(z)$, and $\theta^{*}(z)$ have no common roots and are different from zero. $\phi^{+}(z)$ is a $r$-th order polynomial, $\phi^{*}(z)$ is a $s$-th order polynomial, $\theta^{+}(z)$ is a $r'$-th order polynomial, and $\theta^{*}(z)$ is a $s'$-th polynomial. The order of the polynomials $\phi(z)$ and $\theta(z)$ are $p=r+s$ and $q=r'+s'$. The polynomials $\phi^{+}(z)$ and $\theta^{+}(z)$ have their zeros outside the unit circle. In contrast, the polynomials $\phi^{*}(z)$, and $\theta^{*}(z)$ have their zeros inside the unit circle, that is

\begin{equation}
\label{cond3}
\begin{split}
\phi^{+}(z)\neq0\;\; \textrm{for}\;\; \lvert z\rvert<1\;\;\;\textrm{and}\;\;\;\theta^{+}(z)\neq0\;\; \textrm{for}\;\; \lvert z\rvert<1,\\
\phi^{*}(z)\neq0\;\; \textrm{for}\;\; \lvert z\rvert>1\;\;\;\textrm{and}\;\;\;\theta^{*}(z)\neq0\;\; \textrm{for}\;\; \lvert z\rvert>1.
\end{split}
\end{equation}
\\
In Equation \ref{cond3}, we exclude the unit root case for convenience.
Note that the noncausal component allows a forward-looking representation of $y_t$. Similarly, the noninvertible component allows the forward-looking representation of $\varepsilon_t$. The MARMA($r,s,r',s'$) is represented as 

\begin{equation}
\phi^{+}(z)\phi^{*}(z)y_t=\theta^{+}(z)\theta^{*}(z)\varepsilon_t.
\label{arma}
\end{equation}

Note that our presentation of the model is similar to \cite{lii1992approximate} Equation (1.2), and \cite{fries2021conditional}. Similar representations can be seen for the MA component in \cite{meitz2013maximum,nyholm2019essays}, and for the AR component in \cite{lanne2011noncausal,hecq2016identification}. Under these conditions, the inverses of the lag polynomials are

\begin{equation}
\begin{split}
    \phi^{+}(z)^{-1}=\sum_{j=0}^{\infty}\alpha_jz^{j}\;,\;\;\; \phi^{*}(z)^{-1}=\sum_{j=s}^{\infty}\beta_jz^{-j},\\
    \theta^{+}(z)^{-1}=\sum_{j=0}^{\infty}\alpha'_jz^{j}\;,\;\;\;  \theta^{*}(z)^{-1}=\sum_{j=s'}^{\infty}\beta'_jz^{-j}.
    \end{split}
\end{equation}

The random variables $z_t$ are assumed to have a density function $f_{\sigma}(z)=\frac{1}{\sigma}f(\frac{z}{\sigma})$. Given these representations, the MARMA($r,s,r',s'$) model has a unique and stationary solution, admitting a two-sided representation, an MA($\infty$), and an AR($\infty$), see Theorem 3.1.3 in \cite{brockwell2009time}. The MA($\infty$) representation as  

\begin{equation}\label{doublesum1}
        y_t=\sum_{j=-\infty}^{\infty}\Psi_j\varepsilon_{t+j}, \;\;\textrm{where}\;\;\;\frac{\theta^{+}(z)\theta^{*}(z)}{\phi^{+}(z)\phi^{*}(z)}=\sum_{j=-\infty}^{\infty}\Psi_jz^j=\Psi(z), 
\end{equation}
where $\Psi_j$ are the coefficient of $z^j$ in the Laurent expansion of $\frac{\theta^{+}(z)\theta^{*}(z)}{\phi^{+}(z)\phi^{*}(z)}\overset{\mathrm{def}}{=}\Psi(z)$. Note that $\Psi(z)$ is absolute convergent in the annulus $D=\left\{z:\delta<\lvert z\rvert<\delta^{-1}\right\}, \delta<1$, see Example 1, in \cite{breidt1992time}. Also, the coefficients $\Psi_j$ geometrically decay to zero as $j\rightarrow\infty$. Similarly, $\varepsilon_t$ can be expressed in terms of $y_t$ in an AR($\infty$) representation as

\begin{equation}\label{doublesum2}
    \varepsilon_t=\sum^{\infty}_{j=-\infty}\Psi^{*}_jy_{t+j},\;\;\textrm{where}\;\;\;\frac{\phi^{+}(z)\phi^{*}(z)}{\theta^{+}(z)\theta^{*}(z)}=\sum_{j=-\infty}^{\infty}\Psi^{*}_jz^j=\Psi^{*}(z),   
\end{equation}
where $\Psi^{*}_j$ is the coefficient of $z^j$ in the Laurent expansion of $\frac{\phi^{+}(z)\phi^{*}(z)}{\theta^{+}(z)\theta^{*}(z)}\overset{\mathrm{def}}{=}\Psi^{*}(z)$. $\Psi^{*}(z)$ is absolute convergent in some annulus $D=\left\{z:\delta<\lvert z\rvert<\delta^{-1}\right\}, \delta<1$. The coefficients $\Psi^{*}_j$ geometrically decay to zero as $j\rightarrow\infty$. Note that from the MARMA model can emerge the causal and noncausal AR model, the invertible and noninvertible MA, and the mixed causal-noncausal and mixed invertible-noninvertible, as follows in Remark 1.
\\\\
\textbf{Remark 1}. Particular cases of the MARMA($r,s,r',s'$) model are
\begin{enumerate}
    \item If $\phi^{*}(z)\equiv\theta^{+}(z)\equiv\theta^{*}(z)\equiv1$, the resulting model is a purely causal MAR($r,0$).
   \item If $\phi^{+}(z)\equiv\theta^{+}(z)\equiv\theta^{*}(z)\equiv1$, the resulting model is a purely noncausal MAR($0,s$).

   \item If $\phi^{+}(z)\equiv\phi^{*}(z)\equiv\theta^{*}(z)\equiv1$, the resulting model is an invertible MA($r',0$).

   \item If $\phi^{+}(z)\equiv\phi^{*}(z)\equiv\theta^{+}(z)\equiv1$, the resulting model is a noninvertible  MA($0,s'$).

   \item If $\theta^{+}(z)\equiv\theta^{*}(z)\equiv1$, the resulting model is a mixed causal-noncausal MAR($r,s$).

   \item If $\phi^{+}(z)\equiv\phi^{*}(z)\equiv1$, the resulting model is a mixed invertible-noninvertible MMA($r',s'$).
\end{enumerate}
\vskip 0.3cm
For the estimation of the MARMA($r,s,r',s'$) model, it is convenient to denote the set of parameters as

\begin{equation}
    \boldsymbol{\vartheta}=(\phi^{+}_1,...,\phi^{+}_r,\phi^{*}_1,...,\phi^{*}_s,\theta^{+}_1,...,\theta^{+}_{r'},\theta^{*}_1,...,\theta^{*}_{s'}).
\end{equation}

\subsection{Spectral representation of the MARMA($r,s,r',s'$) process}

This section defines the spectrum for the MARMA($r,s,r',s'$) process. We provide the theory and the intuition to decompose the MARMA process into periodic components that appear in proportion to their variance (second-order cumulant) and higher-order moments/cumulants in the case of the bispectrum. Let us assume $y_t$ be a stationary process with an autocovariance function 

\begin{equation}
    \kappa_2(j)=\mathbb{E}\left[y_t y_{t-j}\right],\;\; \textrm{for}\;\;\ j=\left\{0,\pm 1, \pm2, \pm 3, ... \right\}.
\end{equation}

The autocovariance function is absolute summable  $\sum^{\infty}_{-\infty}\lvert \kappa_2(j) \rvert <\infty$. Therefore, there exists a unique monotonically increasing function $S_2(\omega)$, called the spectrum, with $S_2(-\infty)=S_2(-1/2)=0$, and $S_2(\infty)=S_2(1/2)=\kappa_2(0)$, as

\begin{equation}\label{autocov}
    \kappa_2(j)=\int_{-\pi}^{\pi}S_2(\omega)e^{-ij\omega}d\omega, 
\end{equation}
where $\omega=2\pi/T$ is the Fourier frequency, and $i$ is the imaginary number. The autocovariance considers linear relations of the data in two time periods, known as second-order dependence, also that $\kappa_{2_j}=\kappa_{2_{j-1}}...=\kappa_{2_{j-n}}=\kappa_{2_{j+1}}...=\kappa_{2_{j+n}}$. In any of these representations, the autocovariance functions remain unchanged. We obtain the spectrum by the inverse transform of Equation \ref{autocov}, as

\begin{equation}
    S_2(\omega_j)=\frac{1}{2\pi}\sum_{j=-\infty}^{\infty} \kappa_2(j) e^{-ij\omega}.
\end{equation}

The spectrum and the autocovariance function contain the same information, although expressed in different ways. The autocovariance expresses the dependence between leads or lags, and the spectrum expresses it in cycles, determined by the Fourier frequencies. Of particular interest is the spectrum of the error sequence $\varepsilon_t$. Since its autocovariance function is 

\begin{equation}
    \kappa^{*}_2(j)=\begin{cases}\kappa_2^{*} & j = 0\\0 & j \neq 0\end{cases}.
\end{equation}

The spectrum of $\varepsilon_t$ is $S^\varepsilon_{2}=\kappa_2^{*}$, that is a constant value for all frequencies. 

When $y_t$ is non-Gaussian, its probability structure depends not only on its autocovariance function but also on the higher-order moments.  We specify that the first three moments are equivalent to the first three cumulants, which is not valid for cumulants of order greater than three. In particular, the third cumulant contains the relations of the time series at three points in time. Hence it is known as the triple correlation function or the bicovariance function, see \cite{bartelt1984phase}. The third order cumulant is

\begin{equation}
    \kappa_3(j,l)=\mathbb{E}\left[y_t y_{t-j}y_{t-l}\right],\;\; \textrm{for}\;\;\ j,l=\left\{0,\pm 1, \pm2, \pm 3, ... \right\}.
\end{equation}

Note that $\kappa_3(j,l)$ is equal to zero for the Gaussian distribution. $\kappa_3(j,l)$ satisfies the following symmetries

\begin{equation} \label{cumu}
 \kappa_3(j,l) = \kappa_3(l,j)=\kappa_3(-j,l-j)=\kappa_3(l-j,-j)=\kappa_3(j-l,-l)=\kappa_3(-l,j-l).
\end{equation}

Like the spectrum, the bispectrum is the frequency domain tool for representing non-Gaussian relations. We define the spectrum in terms of the bicovariance function as

\begin{equation}\label{bicov}
    \kappa_3(j,l)=\int_{-\pi}^{\pi}\int_{-\pi}^{\pi}S_3(\omega_1,\omega_2)e^{-i(j\omega_1+l\omega_2)}d\omega_1d\omega_2. 
\end{equation}
We obtain the bispectrum by the inverse transform of Equation \ref{bicov}, as

\begin{equation}
    S_3(\omega_1,\omega_2)=\frac{1}{(2\pi)^2}\sum_{j=-\infty}^{\infty}\sum_{l=-\infty}^{\infty} \kappa_3(j,l) e^{-i(j\omega_1+l\omega_2)}.
\end{equation}

For errors sequence $\varepsilon_t$, the bicovariance function is 

\begin{equation}
    \kappa^{*}_3(j,l)=\mathbb{E}\left[\varepsilon_t \varepsilon_{t-j}\varepsilon_{t-l}\right],
\end{equation}

Thus, the bispectrum of $\varepsilon_t$ is $S^{\varepsilon}_{3}=\kappa^{*}_3(j,l)$. Using this information, we can formulate the spectrum of the MARMA($r,s,r',s'$) model. The linear filter from Equation \ref{doublesum1} is the convolution of the error sequence $\varepsilon_t$ into $y_t$. Then, we can define the frequency impulse response function or transfer function as

\begin{equation}\label{tf}
    \psi( \boldsymbol{\vartheta},\omega)=\sum_{j=-\infty}^{\infty}\Psi_je^{-ij\omega}=\frac{(1+\theta^{+}_1z+...+\theta^{+}_{r'}z^{r'})(1+\theta^{*}_1z+...+\theta^{*}_{s'}z^{s'})}{(1-\phi^{+}_1z-...-\phi^{+}_{r}z^r)(1-\phi^{*}_1z-...-\phi^{*}_sz^s)},
\end{equation}
where $z=e^{-i\omega}$. Note that in \cite{hecq2022spectral} the numerator of the Equation \ref{tf} is equal to one, since they ignore the MA component. In this way, using the spectral density of $\varepsilon_t$, it is possible to deduce the spectral density of $y_t$ as

\begin{equation}
S_2(\boldsymbol{\vartheta},\omega)=\frac{\kappa_2^{*}}{2\pi}\psi(\boldsymbol{\vartheta},\omega)\overline{\psi(\boldsymbol{\vartheta},\omega)}=\frac{\kappa_2^{*}}{2\pi}\lvert \psi(\boldsymbol{\vartheta},\omega)\rvert^2,   
\end{equation}
where $\overline{\psi(\boldsymbol{\vartheta},\omega)}$, is the conjugate of the transfer function. In the same fashion, the bispectrum of $y_t$ is

\begin{equation}
S_3(\boldsymbol{\vartheta},\omega_1,\omega_2)=\frac{\kappa_3^{*}}{4\pi^2}\psi(\boldsymbol{\vartheta},\omega_1)\psi(\boldsymbol{\vartheta},\omega_2)\overline{\psi(\boldsymbol{\vartheta},-\omega_1-\omega_2)},   
\end{equation}
where $\overline{\psi(\boldsymbol{\vartheta},-\omega_1-\omega_2)}$ is the complex conjugate of the transfer function, evaluated at the sum of frequencies $\omega_1$ and $\omega_2$.

\section{Spectral estimation}
\subsection{Spectrum and bispectrum}
The nonparametric estimation of the spectrum and bispectrum for autoregressive processes is usually based on methods employing the discrete Fourier transform (DFT)

\begin{equation}\label{dft1}
    d_T(\omega)=\sum_{t=1}^{T}y_te^{-it\omega}.
\end{equation}

The most popular estimator for the spectrum is the periodogram. For a summary of estimates of the spectrum, see \cite{kay1981spectrum}. The periodogram is obtained as the modulus of the output values from the DFT in Equation \ref{dft1},
performed directly on the time series. The periodogram has been widely used in spectral analysis and autoregressive modeling, see \cite{brillinger1975time, rosenblatt1965stationary,alekseev1996asymptotic,bartlett1950periodogram,brillinger1967asymptotic}. The periodogram is defined as

\begin{equation}
    I_2(\omega)=\frac{1}{2\pi T} d_T(\omega)\overline{d_T(\omega)}.
\end{equation}

Similar to the periodogram, taking into account relations between three frequencies (bifrequencies), the biperiodogram is the bispectrum estimator, defined as

\begin{equation}
    I_3=(\omega_1,\omega_2)=\frac{1}{(2\pi)^2 T}d_T(\omega_1)d_T(\omega_2)\overline{d_T(-\omega_1-\omega_2)},
\end{equation}
where $\overline{d_T(-\omega_1-\omega_2)}$, represents the complex conjugate of the Fourier transform of the sum of two frequencies, accounting for the asymmetries present in the non-Gaussian data. Note that the biperiodogram is a complex quantity due to its last factor of the sum of frequencies. The periodogram and biperiodogram are asymptotically unbiased estimators of the spectrum and bispectrum but inconsistent. For details see \cite{brillinger1975time, rosenblatt1965stationary,alekseev1996asymptotic}. Some properties of the periodogram are developed in \cite{brillinger1967asymptotic}; in particular, the mean, the standard deviation, and its correlation between frequencies.

We employ the estimation function proposed by \cite{velasco2018frequency}. We first perform a preliminary Gaussian likelihood estimation and obtain $\overline{\boldsymbol{\vartheta}}$, assuming the model is causal and invertible. We evaluate the second order estimates in the transfer functions $\psi(\overline{\boldsymbol{\vartheta}},\omega)$, $\psi(\overline{\boldsymbol{\vartheta}},-\omega_i-\omega_j)$ and the second order cumulant $\overline{\kappa}_2$. This normalization by preliminary estimates has no effect on the identification of the parameters since they are invariant to any inversion of the polynomial roots and significantly simplifies the estimation compared with \cite{leonenko1998spectral}. The estimation function is
\\

\begin{equation}
\label{est_func}
R_T(\boldsymbol{\vartheta}) = A_{2T}\sum_{j=1}^{T-1}\left(\frac{ I_2(\omega_j)-S^{*}_2(\boldsymbol{\vartheta},\omega_j)}{ \psi(\overline{\boldsymbol{\vartheta}},\omega_j)\psi(\overline{\boldsymbol{\vartheta}},-\omega_j)}\right)^2+A_{3T}\sum_{j=1}^{T-1}\sum_{i=1}^{T-1}\frac{\lvert I_3(\omega_j,\omega_i)-S^{*}_3(\boldsymbol{\vartheta},\omega_j,\omega_{i}) \rvert^2}{ \psi(\overline{\boldsymbol{\vartheta}},\omega_j)\psi(\overline{\boldsymbol{\vartheta}},\omega_i)\psi(\overline{\boldsymbol{\vartheta}},-\omega_j-\omega_i)},    
\end{equation}
\\
where $A_{2T}=m(2\pi)^2/(4\overline{\kappa_2}^2T)$ and $A_{3T}=n(2\pi)^4/(6\overline{\kappa_2}^3T^2)$. We calculate the second order cumulant by $\overline{k}_{2}=2\pi T^{-1}\sum_{j=1}^{T-1}I_2(\omega_j)/(\psi(\overline{\boldsymbol{\vartheta}},\omega_j)\psi(\overline{\boldsymbol{\vartheta}},-\omega_j))$. The weights in our approach are arbitrary set as $m=0.5$ and $n=0.5$. Nevertheless, in \cite{velasco2018frequency}, the weights are optimally selected depending on the magnitude of the high-order cumulants, increasing the efficiency. However, they turn out to be unintuitive and could increase the complexity of the estimation. Considering Theorem 2 in \cite{velasco2018frequency}, we include in the spectrum and the bispectrum a consistent estimator of the standardized cumulant of orders two and three, respectively, that is

\begin{equation}
    k^{*}_2(\boldsymbol{\vartheta})=\frac{2\pi}{T}\sum_{j=1}^{T-1}\frac{I_2(\omega_j)}{\psi(\boldsymbol{\vartheta},\omega_j)\psi(\boldsymbol{\vartheta},-\omega_j)},\;\; \;\;\;\;k^{*}_3(\boldsymbol{\vartheta})=\frac{4\pi^2}{T^2} \sum_{j=1}^{T-1}\sum_{i=1}^{T-1}Re\left(\frac{I_3(\omega_j,\omega_i)}{\psi(\boldsymbol{\vartheta},\omega_j)\psi(\boldsymbol{\vartheta},\omega_i)\psi(\boldsymbol{\vartheta},-\omega_j-\omega_i)}\right).
\end{equation}

For a correct estimation and identification, the parameters and cumulants must be jointly estimated. Therefore, we use the modified spectrum and bispectrum, where the cumulants are dependent on $\boldsymbol{\vartheta}$. The modified spectrum and bispectrum are

\begin{equation}
S^{*}_2(\boldsymbol{\vartheta},\omega_j)=S_2(\boldsymbol{\vartheta},k^{*}_2(\boldsymbol{\vartheta}),\omega_j);\;\;\;S^{*}_3(\boldsymbol{\vartheta},\omega_j,\omega_i)=S_3(\boldsymbol{\vartheta},k^{*}_3(\boldsymbol{\vartheta}),\omega_j,\omega_i).
\end{equation}
\\
Notice that he first term in $R_T(\boldsymbol{\vartheta})$ is equivalent to
\cite{whittle1953estimation} or the Gaussian-likelihood. This consistency and asymptotic normality of the estimator is developed in \cite{terdik1999bilinear,velasco2018frequency}. The set of estimated parameters $\hat{\boldsymbol{\vartheta}}$ is obtained by minimizing $R_T$

\begin{equation}
\hat{\boldsymbol{\vartheta}}=\argmin_{\boldsymbol{\vartheta} \in \Theta} \: R_T(\boldsymbol{\vartheta}),
    \end{equation}
where the parametric space $\Theta$ of $\boldsymbol{\vartheta}$ is assumed to be a subset of an Euclidean space  $\mathbb{R}^{r+s+r'+s'+2}$. The global minimum is achieved if $\boldsymbol{\hat{\vartheta}} \in \Theta$, and if $ R_T(\boldsymbol{\hat{\vartheta}})\leq R_T(\boldsymbol{\vartheta}), \forall \boldsymbol{\vartheta} \in \Theta$. Note that the location of the roots outside and inside the unit circle is directly made in the transfer function in Equation \ref{tf}. So far, we only talked about model identifications and parameter estimations. The way the computation of the standard errors is done can be found in Appendix A.

\subsection{Multimodality and estimation methods}

Multimodality in the estimation functions poses significant challenges when identifying and estimating noncausal and mixed models. Specifically, the presence of multiple local minima, which may closely resemble the global minimum yet reside in distinct locations, often complicates the estimation process. Conventional estimation techniques grounded in the second moment might falter in discerning between these minima. Consequently, such ambiguities can incorrectly identify noncausal models as causal and vice versa.

It's paramount to understand that multimodality is an intrinsic trait of noncausal and mixed models and not only a byproduct of the estimation methods themselves. Manifestations of multimodality are evident in MLE undertaken in both the time domain \cite{hecq2016identification,bec2020mixed,kindop2021ubiquitous}. A few potential solutions to the multimodality problem have emerged, e.g., \cite{cubadda2023optimization} proposed to use the Simulated Annealing Algorithm. Similarly, \cite{hecq2022spectral} finds that the multimodality amplifies as the data distribution inches closer to normality and as the disparity between the roots of causal and noncausal models enlarges. Notably, multimodality can endure even for large samples.

In our context, the spectral estimation function, $R_T(\boldsymbol{\vartheta})$, presents intricate complexities. Pursuing an optimal solution via mathematical derivation becomes non-feasible given the function's topology in the parameter space $\Theta$. This complexity is epitomized by pronounced peaks in proximity to the unit circle and steep descents as the parameter nears zero. To circumvent the challenges of multimodality and achieve precise estimations, we incorporate two optimization strategies: the Gradient Descent, and the Genetic Algorithm.

\subsubsection{Gradient Descendent Methods}

Conventional gradient descent algorithms, such as Newton-Raphson and Nelder-Mead, may exhibit limitations when the initial values provided for optimization significantly deviate from the true values. Initial values are usually unimportant asymptotically but important in finite samples. Consequently, there is a high likelihood of becoming trapped in a local minimum, leading to inaccurate identification and estimation of potential noncausal or mixed models. To address this concern, \cite{hecq2022spectral} employ the computational algorithm introduced in Section 3.2 of to select an initial values that closely approximates the true set of parameters using the flipping root approach. This process can become intricate as the number of parameters increases, owing to the factorization of the characteristic function of the polynomials. Furthermore, the number of potential solutions escalates with an increasing number of parameters. To address this challenge, we use a Genetic Algorithm method.

\subsubsection{Genetic Algorithm}

Genetic Algorithms are a type of optimization algorithm that draws inspiration from the process of natural selection and genetics. They are widely used to address complex optimization problems. The main objective of the Genetic Algorithm is to mimic the process of evolution to iteratively search for the optimal solution in the parameter space $\Theta$. The initial stage involves creating a population of potential solutions to the problem, representing each solution as a "chromosome" in the population. Typically, these chromosomes are encoded as binary digits (genes) strings. In the case of our study, multiple parameters that could lead to the global minimum of the estimation function $R_T(\boldsymbol{\vartheta})$ are randomly generated. Following this, all possible parameters are evaluated, and the ones that show better results are selected, resembling the survival process of living organisms. The best results are then reused in the crossover and mutation process, introducing small random changes to the parameter set in the search for the global minimum. The algorithm continues the process of selection and reproduction for a fixed number of generations or until a termination condition is met, such as the estimation function showing no significant changes. Through the iterative application of the algorithm, the entire parameter space is explored, favoring solutions with higher fitness values over time. Thus, the initial values for the estimation become irrelevant.

\subsection{Identification strategy}
To achieve the correct identification of the MARMA model we follow three steps. 

\begin{enumerate}
    \item We perform the Jarque-Bera to $y_t$ (or on the residuals from the pseudo-causal model). If we reject the null hypothesis of normality, we perform steps 2 and 3. Otherwise, if the null hypothesis is not rejected, we can only estimate causal and invertible ARMA models. 
    
    \item We determine the dynamics of $y_t$ using the ARMA($p,q$) model assuming causality and invertibility. We use the AIC aor BIC to select the order of $p$ and $q$. In this way, there are $2^{p+q}$ combinations of roots that could have the same second-order probability structure. However, if $y_t$ is non-Gaussian, only one model corresponds to the data generating process (DGP) of the data and can be found using $R_T(\boldsymbol{\vartheta})$.

    \item We estimate all the combinations of MARMA($r,s,r',s'$) models that satisfy the order $p=r+s$ and $q=r'+s'$ using the Genetic Algorithm. The model that best fits the data emerges from the global minimum of the estimation function $R_T(\boldsymbol{\vartheta})$ of all possible models. 
    \end{enumerate}
\vskip 0.25cm
\textbf{Example 1}: Let us assume in the first step, we reject the null of Gaussianity. Then, it emerges for instance, that in the second step, we obtain the pseudo causal and invertible ARMA(2,2) model using BIC, i.e. $p=2$ and $q=2$. There exist nine possible MARMA($r,s,r',s'$) models to be estimated

\begin{table}[H]
\tiny
\centering
\caption{All possible MARMA($r,s,r',s'$) given the order $p=2$ and $q=2$}
\vskip 0.2cm
\label{marma_mod}
\begin{tabular}{|l|c|c|c|c|}
\hline
\textbf{Model/Order} & \textbf{$r$} & \textbf{$s$} & \textbf{$r'$} & \textbf{$s'$} \\ \hline
\multicolumn{1}{|c|}{MARMA(2,0,2,0)} & 2 & 0 & 2 & 0 \\ \hline
\multicolumn{1}{|c|}{MARMA(1,1,2,0)} & 1 & 1 & 2 & 0 \\ \hline
MARMA(0,2,2,0) & 0 & 2 & 2 & 0 \\ \hline
MARMA(2,0,1,1) & 2 & 0 & 1 & 1 \\ \hline
MARMA(1,1,1,1) & 1 & 1 & 1 & 1 \\ \hline
MARMA(0,2,1,1) & 0 & 2 & 1 & 1 \\ \hline
MARMA(2,0,0,2) & 2 & 0 & 0 & 2 \\ \hline
MARMA(1,1,0,2) & 1 & 1 & 0 & 2 \\ \hline
MARMA(0,2,0,2) & 0 & 2 & 0 & 2 \\ \hline
\end{tabular}
\end{table}

Under non-Gaussianity, and after estimating $R_T(\boldsymbol{\hat{\vartheta}})$ we have the following possible minima
\begin{enumerate}
    \item There exist at least eight local minima, such as $\boldsymbol{\hat{\vartheta}} \in \Theta$ and $\exists\epsilon>0$ subject to $R_T(\boldsymbol{\hat{\vartheta}})\leq R_T(\boldsymbol{\vartheta}), \forall \boldsymbol{\vartheta} \in B(\boldsymbol{\hat{\vartheta}},\epsilon)\; \cap\; \Theta$. where $B(\boldsymbol{\hat{\vartheta}},\epsilon):=\left\{\boldsymbol{\vartheta}\lvert\;\ \lvert\lvert\boldsymbol{\vartheta}-\boldsymbol{\hat{\vartheta}}\rvert\rvert\leq\epsilon \right\}$.
    \item There exists one global minimum, such as $\boldsymbol{\hat{\vartheta}} \in \Theta$, and $ R_T(\boldsymbol{\hat{\vartheta}})\leq R_T(\boldsymbol{\vartheta}), \forall \boldsymbol{\vartheta} \in \Theta$.
\end{enumerate}

Consequently, we estimate the nine models and then select the global minimum.

\subsection{Estimated Residuals}
To diagnose the adequacy of the estimated time-series model and do traditional inference, we first obtain the residuals in the frequency domain as follows
\begin{equation}
d_{\hat{\varepsilon}}(\omega)=d_T(\omega)\frac{(1-\widehat{\phi}^{+}_1z-...-\widehat{\phi}^{+}_rz^r)(1-\widehat{\phi}^{*}_1z-...-\widehat{\phi}^{*}_sz^s)}{(1+\widehat{\theta}^{+}_1z+...+\widehat{\theta}^{+}_{r'}z^{r'})(1+\widehat{\theta}^{*}_1z+...+\widehat{\theta}^{*}_{s'}z^{s'})}.
\end{equation}

Time domain residuals $\widehat{\varepsilon_t}$ are obtained by the real part of the inverse of the Fourier transform of the frequency domain residuals, as

\begin{equation}\label{dft2}
    \widehat{\varepsilon}_t=Re\left(\frac{1}{T}\sum_{j=0}^{T} d_{\hat{\varepsilon}}(\omega)e^{ij\omega}\right).
\end{equation}

Note that the methods performed directly in the time domain (see, \cite{hecq2016identification}) must necessarily lose data due to the inclusion of initial and final conditions in terms of leads of the endogenous variable and future errors. Consequently, they generate a sequence of residuals smaller than the number of data in the time series. On the contrary, with this method, no data is lost since we work directly with the estimated coefficients and the polynomials of Equation \ref{dec}. Estimate residuals will also be used for the computation of standard errors.

\section{Testing for iidness}

We utilize the tests recommended by \cite{dalla2020robust} to assess the iidness of the estimated residuals. Evaluating the i.i.d assumption is important, since the incorrect beliefs about causality or invertibility might produce residuals that aren't independent, especially for orders beyond two. While these residuals might seem like white noise, they won't effectively capture key characteristics of financial returns, such as explosive trends and asymmetric cycles. On the other hand, correctly pinpointing a time series with potential noncausal or noninvertible dynamics leads to error sequences closer to the i.i.d assumption. This diminishes the necessity for models addressing heteroscedasticity, like the GARCH model (refer to \cite{gourieroux2013explosive}, \cite{lanne2013testing}).

\vskip 0.2cm
\textbf{Example 1} 
\textit{Let us illustrate the appearance of conditional  heteroscedastic effects with a noncausal and invertible MARMA (MARMA(0,1,1,0) process) that is estimated as a traditional ARMA model. Let us consider}

\begin{equation}
    (1-\phi^{*}_1z^{-1})y_t=(1+\theta^{+}_1z)\varepsilon_t.
\end{equation}
\textit{that has a unique and stationary representation}

\begin{equation}
    y_t=\sum_{j=-\infty}^{\infty}\Psi_{j}\varepsilon_{t-j},
\end{equation}
\textit{where $\Psi_j$ emerges from the Laurent expansion of $(1-\phi_1^{*}z^{-1})^{-1}(1+\theta z)\overset{\mathrm{def}}{=}\Psi(z)$ and is represented by}

\begin{equation}
    \Psi_j=\begin{cases}0 & j < 0\\-\frac{1}{\phi_1^{*}} & j =1 \\-\phi_1^{*-j}(1+\theta_1^{+}\phi_1^{*}) & j\geq2\end{cases}.
\end{equation}

\textit{Assuming that $\varepsilon_t$ has finite moments up until four, the autocorrelation function of $y_t^2$ is}

\begin{equation}\label{sqr_res}
\textrm{corr}\left(y_t^2,y_{t+k}^2\right)=\frac{(k^{*}_4-3)\sum_{-\infty}^{\infty}\Psi_j^2\Psi_{j+k}^2+2\left(\sum_{-\infty}^{\infty}\Psi_j\Psi_{j+k}\right)^2}{(k^{*}_4-3)\sum_{-\infty}^{\infty}\Psi_j^4+2\left(\sum_{-\infty}^{\infty}\Psi_j^2\right)^2},
\end{equation}
\vskip 0.20cm

where $\kappa_4^{*}$ is the standardized kurtosis as $k_4^{*}=\mathbb{E}(\varepsilon_t^4)/\sigma^4$. For $k>1$ and non-Gaussian data, the correlation between the squares differs from zero. Equation \ref{sqr_res} shows the apparition of heteroscedastic effects when the causal ARMA(1,1) is fitted, and the true DGP is an MARMA(0,1,1,0). 

\vskip 0.2cm
To introduce the tests of \cite{dalla2020robust}, we define the serial correlation, the correlation between absolute values, and the correlation between the squares, respectively as

\begin{equation}
    \rho_{\varepsilon, k}=\textrm{corr}\left(\varepsilon_t,\varepsilon_{t-k}\right),\;\;  \rho_{\lvert\varepsilon,\rvert k}=\textrm{corr}\left(\lvert\varepsilon_t\rvert,\lvert\varepsilon_{t-k}\rvert\right),\;\;  \rho_{\varepsilon^2, k}=\textrm{corr}\left(\varepsilon_t^2,\varepsilon_{t-k}^2\right).
\end{equation}

If $\varepsilon_t$ is i.i.d, then $\rho_{\varepsilon, k}=\rho_{\lvert\varepsilon,\rvert k}=\rho_{\varepsilon^2, k}=0$. Then, testing for iidness reduces to the absence of correlation between the levels $\varepsilon_t$ and the absolute value $\lvert \varepsilon_t\rvert$, and the levels $\varepsilon_t$ and the squares $\varepsilon_t^2$. The first test is produced for individuals lags and cumulative as

\begin{equation}
    J_{\varepsilon,\lvert\varepsilon\rvert,k}=\frac{T^2}{T-k}\left(\widehat{\rho}^2_{\varepsilon,k}+\widehat{\rho}^2_{\lvert \varepsilon \rvert,k}\right),\;\;C_{\varepsilon,\lvert\varepsilon\rvert,k}=\sum_{k=1}^m J_{\varepsilon,\lvert\varepsilon\rvert,k},
\end{equation}
with an individual lag null hypothesis, $H_{0}$:  $\rho_{\varepsilon,k}=0$, $\rho_{\lvert\varepsilon\rvert, k}$ for $k\geq1$, and a cumulative null hypothesis, $H_{0}$:  $\rho_{\varepsilon,k}=0$, $\rho_{\lvert\varepsilon\rvert, k}$ for $k=1,...,m\geq1$. The second test is for the squares, also with individual lag and cumulative statistics

\begin{equation}
    J_{\varepsilon,\varepsilon^2,k}=\frac{T^2}{T-k}\left(\widehat{\rho}^2_{\varepsilon,k}+\widehat{\rho}^2_{ \varepsilon^2,k}\right),\;\;C_{\varepsilon,\varepsilon^2,k}=\sum_{k=1}^m J_{\varepsilon,\varepsilon^2,k},
\end{equation}
with individual lag null hypothesis, $H_{0}$:  $\rho_{\varepsilon,k}=0$, $\rho_{\varepsilon^2, k}$ for $k\geq1$, and cumulative null hypothesis, $H_{0}$:  $\rho_{\varepsilon,k}=0$, $\rho_{\varepsilon^2, k}$ for $k=1,...,m\geq1$. If $\varepsilon_t$ is i.i.d, the asymptotic distributions of the tests are

\begin{equation}
    J_{\varepsilon,\lvert\varepsilon\rvert,k},   J_{\varepsilon,\varepsilon^2,k}\overset{D}{\to}\chi^2_2\;\;\;\; C_{\varepsilon,\lvert\varepsilon\rvert,k},   C_{\varepsilon,\varepsilon^2,k}\overset{D}{\to}\chi^2_{2m}\;\;\;\;\;\;as\;\; T\rightarrow\infty.
\end{equation}

These tests have good power in the presence of dependence, conditional heteroskedasticity, or nonstationarity. Also, the $\chi^2_{2m}$ is a good approximation for the asymptotic distribution of the test for a nonsymmetric $\varepsilon_t$. Similar results are obtained in \cite{mcleod1983diagnostic, wong2005mixed}.

\section{Simulation of MARMA processes}
We propose a method to simulate MARMA processes in the frequency domain. Using the inverse Fourier transform, we can reconstruct the time series $y_t$. Unlike existing methods, i.e., \cite{lanne2011noncausal}, or \cite{hecq2016identification}, we do not use the representation of the series using lags and leads, with the advantage of not having to truncate their values to reach an established order. We use the transfer function in terms of its roots outside or inside the unit circle. 
\\

The steps for simulating an MARMA($r,s,r',s'$) are as follows

\begin{center}
     \begin{figure}[h]
         \centering         
           \caption{Trajectory of a MARMA(1,1,1,1) alpha-stable process}
     \includegraphics[width=0.8 \textwidth]{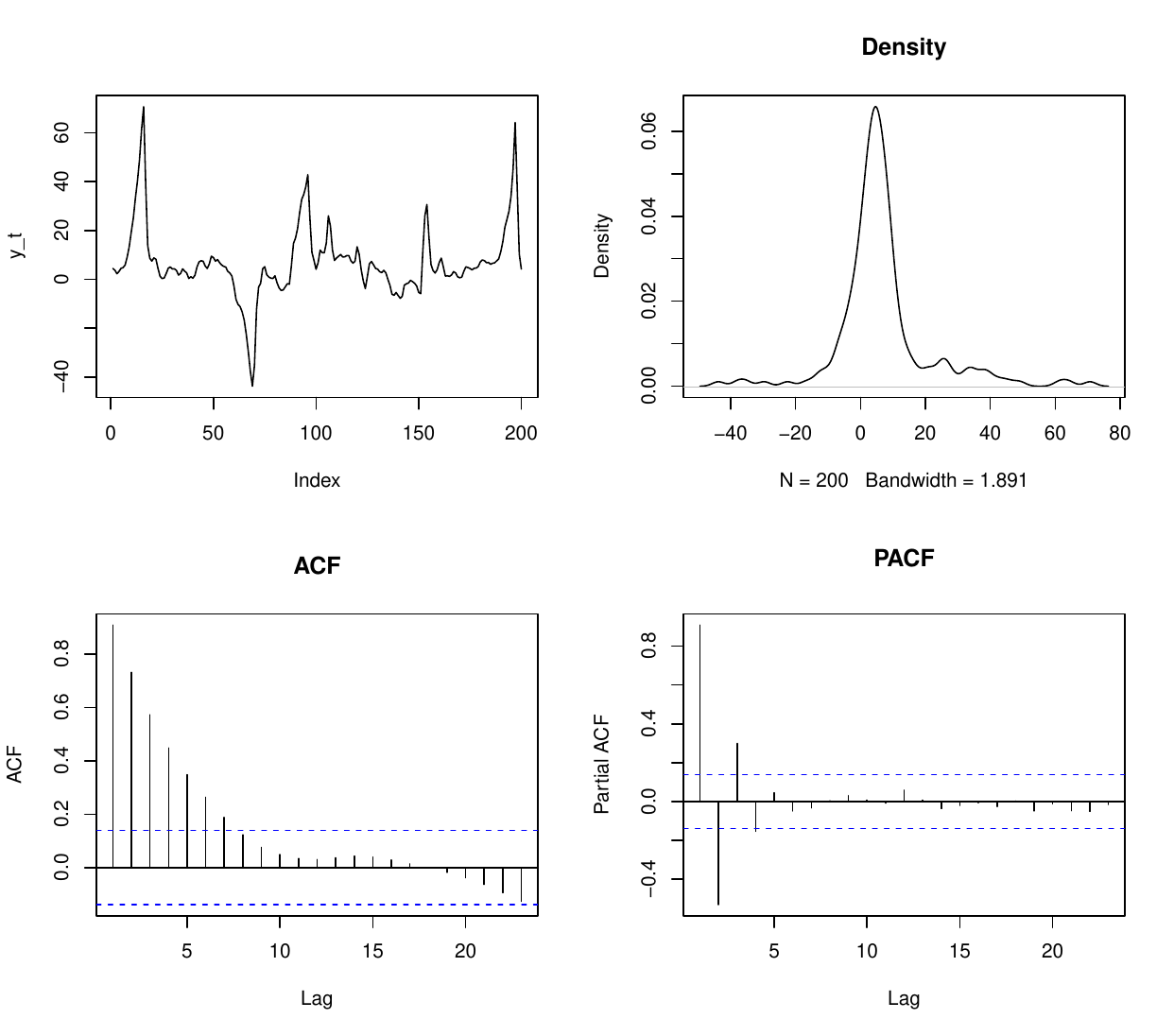}
        \label{marma_1111}
     \end{figure}
 \end{center}

\begin{enumerate}
    \item Simulate $T$ realizations of an i.i.d random error sequence, that is $\left\{\varepsilon_t\right\}_{t=0}^T$, $\varepsilon_t \overset{\mathrm{i.i.d}}{\sim} \mathfrak{G}(\zeta)$, and $\mathfrak{G}$ is a distribution function, and $\zeta$ its parameters.

    \item Perform the DTF of $\varepsilon_t$, that is $d_{\varepsilon}(\omega)=\sum_{t=1}^{T}\varepsilon_te^{-it\omega}$.
    \item Select the parameters and the orders of the MARMA($r,s,r',s'$) process, and create the transfer function.

\[ \psi( \boldsymbol{\vartheta},\omega)=\frac{(1+\theta^{+}_1z+...+\theta^{+}_{r'}z^{r'})(1+\theta^{*}_1z+...+\theta^{*}_{s'}z^{s'})}{(1-\phi^{+}_1z-...-\phi^{+}_rz^{r})(1-\phi^{*}_1z-...-\phi^{*}_sz^s)}.\]

\item Compute the DTF of $y_t$ as

\[d_T(\omega)=d_\varepsilon(\omega)\psi( \boldsymbol{\vartheta},\omega).\]

\item Recover $y_t$ as the real part of the Inverse Fourier Transform of $d_T(\omega)$

\[y_t=Re\left(\frac{1}{T}\sum_{j=0}^{T} d_T(\omega)e^{ij\omega}\right).\]

\end{enumerate}
\vskip 0.5cm
To illustrate our approach we simulate an MARMA($1,1,1,1$) with parameters $\phi_1^{+}=0.7, \phi_1^{*}=-0.2, \theta^{+}_1=0.5$ and $\theta^{*}_1=0.3$. The error sequence is alpha-stable $f_a(x;\alpha,\beta,\eta,\delta)$ with parameters $\alpha=1.5$, $\beta=0.2$, $\eta=1$, $\delta=0$ and $T=200$. The pdf of the alpha-stable is introduced in Section \ref{ss}. In Figure \ref{marma_1111} we plot the trajectories of $y_t$, the empirical density, the ACF, and the PACF. The trajectories of the MARMA(1,1,1,1) processes are characterized by abrupt peaks or bubble phenomena, asymmetric cycles, and some short trend episodes. The ACF and PACF display patterns similar to an ARMA(2,2). 

\begin{center}
     \begin{figure}[h]
         \centering         
           \caption{Real and Imaginary part of the bispectrum and biperiodogram for a MARMA(1,1,1,1) alpha-stable process}
     \includegraphics[width=0.9 \textwidth]{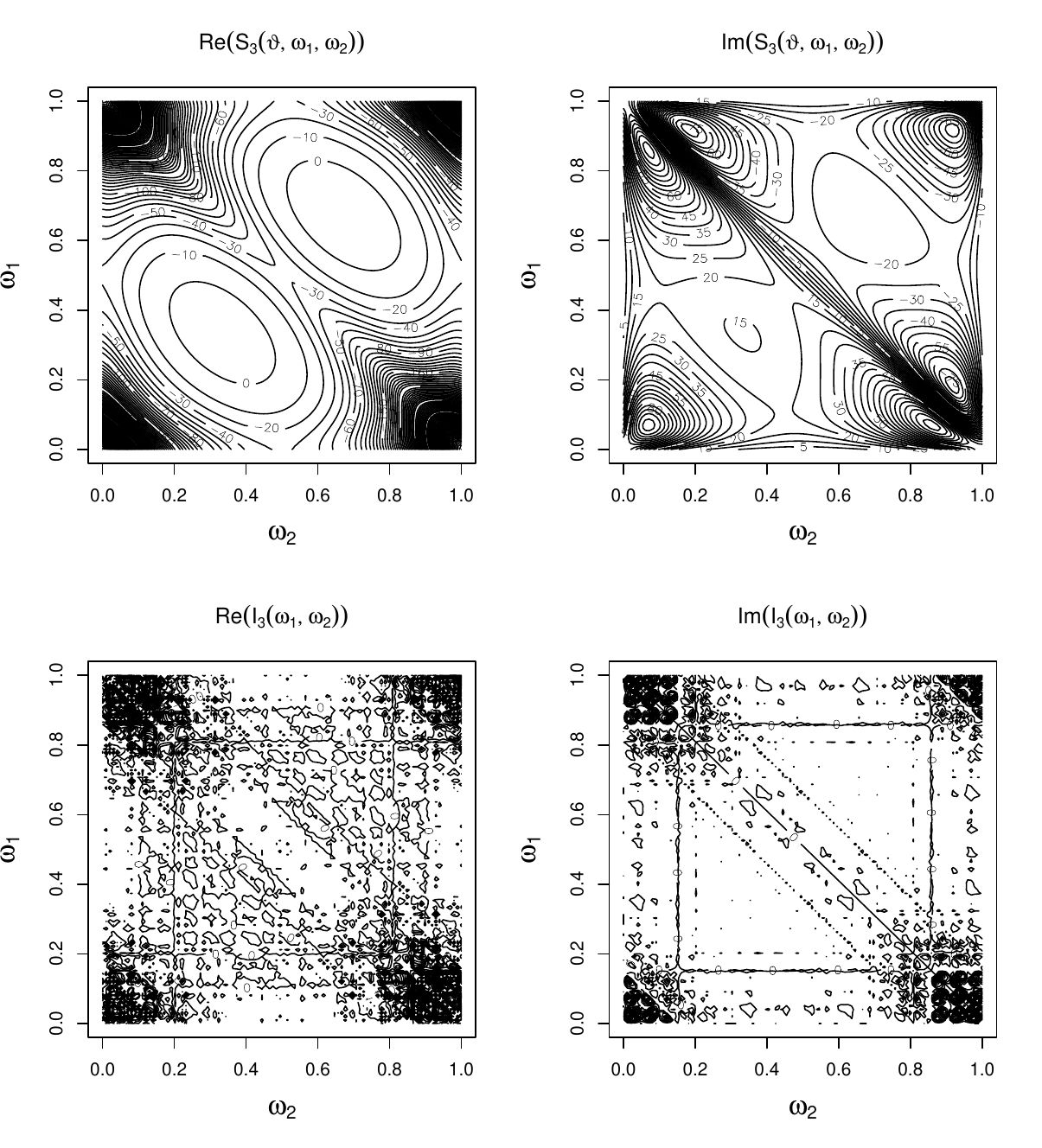}
        \label{marma_spec}
     \end{figure}
 \end{center}

Figure \ref{marma_spec} presents the real and imaginary part of the bispectrum and the biperiodogram of the same MARMA(1,1,1,1) process. Note that the patterns observed in the biperiodogram are replicated by the bispectrum. Also, we observe the same symmetries of the third-order cumulant in Equation \ref{cumu}. The white spaces imply that the relation between these frequencies is null. On the contrary, when it tends to become black, it implies a greater relation. The highest relations occur at the lowest frequencies, $\omega_1$, and $\omega_2$ from 0 to 0.2 and 0.8 to 1. The relation fades as they approach the higher frequencies, from 0.4 to 0.5 and from 0.5 to 0, see \cite{hecq2022spectral} for details.

\section{Simulation study}
\label{ss}
We evaluate the performance of the estimator in finite samples, focusing on the correct identification and estimation of the MARMA($r,s,r',s'$) model. This entails discerning the causal from the noncausal autoregressive components and the invertible from the noninvertible moving average components. It should be noted that \cite{velasco2018frequency} conducted a simulation study for the invertible and nononvertible MA(1) model, utilizing the third and fourth-order spectral densities in the estimation. In contrast, our research centers on the MARMA models with two parameters, employing only the second and third-order spectral densities. 

We focus on three type of models:
\begin{enumerate}
    \item Autoregressive models: MAR(2,0), MAR(0,2), and MAR(1,1).
    \item Moving average models: MA(2,0), MA(0,2), and MMA(1,1).
    \item Combined models: MARMA(1,0,1,0), MARMA(0,1,1,0), MARMA(1,0,0,1), and MARMA(0,1,0,1).
\end{enumerate}

We employ two error sequence for $\varepsilon_t$, that is the alpha-stable and the skew-t distribution, as proposed by \cite{fernandez1998bayesian}. The pdf of the alpha-stable is

\begin{equation}
    f_a(x;\alpha,\beta,\eta,\delta)=exp(ix\delta-\lvert\eta x\rvert^\alpha(1-i\beta sgn(x)(\lvert\eta x\rvert^{1-\alpha}-1)tan(\pi\alpha2^{-1}))),
\end{equation}
where $\alpha\in(0,2)$ is the stability parameter, $\beta \left[-1,1\right]$ is the skewness parameter, $\eta>0$ the scale parameter, and $\delta\in\Re$ is the location parameter. Using the alpha-stable distribution has the benefit of replicating sample features of financial returns, such as leptokurticity and skewness. However, the existence of its moments is linked to the stability parameter ($\alpha$), i.e., the variance, the skewness, and the kurtosis only exist when it is equal to two. An interesting feature of the alpha-stable distribution is its infinite divisibility, i.e., it can be expressed as a sum of i.i.d. variables while preserving its distribution. The pdf of the skew-t is 
\begin{equation}
    t_{skew}(x)=\begin{cases}\frac{2}{\gamma+1/\gamma}t(\gamma x) & for \;\;x < 0\\\frac{2}{\gamma+1/\gamma}t(\frac{x}{\gamma}) &for \;\; x \geq 0\end{cases}, \;\; t(x)=\frac{\Gamma\left(\frac{\nu+1}{2}\right)}{\sqrt{\nu\pi}\Gamma\left(\frac{\nu}{2}\right)}\left(1+\frac{x^2}{\nu}\right)^{-(\nu+1)/2},
\end{equation}
where $\gamma\in\Re$ is the skewing parameter, and $t(x)$ is the pdf of the student's t distribution, where $\nu>2$ is the degrees of freedom. The skew-t distribution can also reproduce heavy-tailed and skewed patterns. Nevertheless, its moments exist as a function of degrees of freedom. For example, the mean exists for $\nu>1$, the variance for $\nu>2$, the skewness for $\nu>3$, the kurtosis for $\nu>4$, and so forth.

For the alpha stable we consider two stability parameters $\alpha=\left\{1.2, 1.8\right\}$, one skewness parameter $\beta=0.25$, and $\eta=1$ and $\delta=0$. For the skew-t, we consider two degrees of freedom $\nu=\left\{2, 5\right\}$ and skewness parameter $\gamma=1.1$. Note that in both cases, one distribution is more leptokurtic than the other, which allows us to make comparisons when it departs from normality to a greater extent. We also consider data lengths $T=\left\{100, 300, 500\right\}$, and a number of simulations $M=1000$. In all cases, we report the mean estimate, the standard deviation, and the number of times the identification was correct (identification rate). Throughout the study, we employed the Genetic Algorithm for the estimation.

\subsection{The MAR($r,s$) DGP}
We first consider the causal MAR(2,0) DGP with parameters $\phi^{+}_1=0.7$ and $\phi^{+}_2=0.2$. Then we estimate three models: a purely causal MAR(2,0), a purely noncausal MAR(0,2), and the mixed MAR(1,1). The identification rate corresponds to the the frequency with which we get the correct MAR(2,0). Accordingly, the mean and standard deviation correspond to the estimated parameters of the correctly identified MAR(2,0) model in the $M$ iterations. Subsequently, we consider as the DGP the MAR(0,2) with parameters $\phi^{*}_1=0.7$ and $\phi^{*}_2=0.2$, and, finally, the MAR(1,1) with parameters $\phi^{+}_1=0.7$ and $\phi^{*}_1=0.2$. The same comparisons are made to establish the identification rate in both cases. We also report the mean and standard deviation of the estimated parameters. The results are presented in Table \ref{ar_mc}. 

\begin{table}[H]
\centering
\tiny
\caption{Identification rate, the mean and standard deviation of estimated parameters of MAR(2,0), MAR(1,1), and MAR(0,2) with errors alpha-stable and skew-t}
\vskip 0.2cm
\label{ar_mc}
\begin{tabular}{cccccccccccccc}
\hline
\multicolumn{1}{l}{} & \multicolumn{1}{l}{} & \multicolumn{12}{c}{\textbf{Panel A: Alpha-stable}} \\ \hline
\multicolumn{1}{l}{} & \multicolumn{1}{l}{} & \multicolumn{4}{c}{MAR(2,0)} & \multicolumn{4}{c}{MAR(1,1)} & \multicolumn{4}{c}{MAR(0,2)} \\ \cline{3-14} 
\multicolumn{1}{l}{Dist} & T & \multicolumn{4}{c}{Identification rate} & \multicolumn{4}{c}{Identification rate} & \multicolumn{4}{c}{Identification rate} \\ \hline
\multirow{3}{*}{$\alpha=1.2$} & 100 & \multicolumn{4}{c}{80\%} & \multicolumn{4}{c}{81\%} & \multicolumn{4}{c}{82\%} \\
 & 300 & \multicolumn{4}{c}{94\%} & \multicolumn{4}{c}{94\%} & \multicolumn{4}{c}{93\%} \\
 & 500 & \multicolumn{4}{c}{96\%} & \multicolumn{4}{c}{96\%} & \multicolumn{4}{c}{98\%} \\
\multirow{3}{*}{$\alpha=1.8$} & 100 & \multicolumn{4}{c}{47\%} & \multicolumn{4}{c}{55\%} & \multicolumn{4}{c}{50\%} \\
 & 300 & \multicolumn{4}{c}{75\%} & \multicolumn{4}{c}{72\%} & \multicolumn{4}{c}{77\%} \\
 & 500 & \multicolumn{4}{c}{84\%} & \multicolumn{4}{c}{82\%} & \multicolumn{4}{c}{86\%} \\ \hline
\multicolumn{1}{l}{} & \multicolumn{1}{l}{} & \multicolumn{2}{c}{$\phi^{+}_1=0.7$} & \multicolumn{2}{c|}{$\phi^{+}_2=0.2$} & \multicolumn{2}{c}{$\phi^{+}_1=0.7$} & \multicolumn{2}{c|}{$\phi^{*}_1=0.2$} & \multicolumn{2}{c}{$\phi^{*}_1=0.7$} & \multicolumn{2}{c}{$\phi^{*}_2=0.2$} \\ \cline{3-14} 
\multicolumn{1}{l}{Dist} & T & Mean & Std & Mean & \multicolumn{1}{c|}{Std} & Mean & Std & Mean & \multicolumn{1}{c|}{Std} & Mean & Std & Mean & Std \\ \hline
\multirow{3}{*}{$\alpha=1.2$} & 100 & 0.721 & 0.135 & 0.153 & \multicolumn{1}{c|}{0.118} & 0.670 & 0.123 & 0.203 & \multicolumn{1}{c|}{0.136} & 0.713 & 0.117 & 0.172 & 0.120 \\
 & \multicolumn{1}{l}{300} & 0.694 & 0.070 & 0.196 & \multicolumn{1}{c|}{0.089} & 0.695 & 0.057 & 0.215 & \multicolumn{1}{c|}{0.079} & 0.703 & 0.081 & 0.182 & 0.082 \\
 & \multicolumn{1}{l}{500} & 0.695 & 0.065 & 0.198 & \multicolumn{1}{c|}{0.070} & 0.702 & 0.025 & 0.196 & \multicolumn{1}{c|}{0.046} & 0.701 & 0.035 & 0.205 & 0.047 \\
\multirow{3}{*}{$\alpha=1.8$} & 100 & 0.734 & 0.130 & 0.127 & \multicolumn{1}{c|}{0.140} & 0.665 & 0.132 & 0.211 & \multicolumn{1}{c|}{0.165} & 0.736 & 0.118 & 0.150 & 0.128 \\
 & 300 & 0.707 & 0.087 & 0.182 & \multicolumn{1}{c|}{0.099} & 0.678 & 0.113 & 0.209 & \multicolumn{1}{c|}{0.113} & 0.697 & 0.087 & 0.178 & 0.104 \\
 & 500 & 0.702 & 0.076 & 0.193 & \multicolumn{1}{c|}{0.081} & 0.697 & 0.088 & 0.203 & \multicolumn{1}{c|}{0.092} & 0.717 & 0.081 & 0.193 & 0.097 \\ \hline
\multicolumn{1}{l}{} & \multicolumn{1}{l}{} & \multicolumn{1}{l}{} & \multicolumn{1}{l}{} & \multicolumn{1}{l}{} & \multicolumn{1}{l}{} & \multicolumn{1}{l}{} & \multicolumn{1}{l}{} & \multicolumn{1}{l}{} & \multicolumn{1}{l}{} & \multicolumn{1}{l}{} & \multicolumn{1}{l}{} & \multicolumn{1}{l}{} & \multicolumn{1}{l}{} \\ \hline
\multicolumn{1}{l}{} & \multicolumn{1}{l}{} & \multicolumn{12}{c}{\textbf{Panel B: Skew t}} \\ \hline
\multicolumn{1}{l}{} & \multicolumn{1}{l}{} & \multicolumn{4}{c}{MAR(2,0)} & \multicolumn{4}{c}{MAR(1,1)} & \multicolumn{4}{c}{MAR(0,2)} \\ \cline{3-14} 
\multicolumn{1}{l}{Dist} & T & \multicolumn{4}{c}{Identification rate} & \multicolumn{4}{c}{Identification rate} & \multicolumn{4}{c}{Identification rate} \\ \hline
\multirow{3}{*}{$\nu=2$} & 100 & \multicolumn{4}{c}{68\%} & \multicolumn{4}{c}{68\%} & \multicolumn{4}{c}{73\%} \\
 & 300 & \multicolumn{4}{c}{91\%} & \multicolumn{4}{c}{74\%} & \multicolumn{4}{c}{93\%} \\
 & 500 & \multicolumn{4}{c}{96\%} & \multicolumn{4}{c}{90\%} & \multicolumn{4}{c}{95\%} \\
\multirow{3}{*}{$\nu=5$} & 100 & \multicolumn{4}{c}{54\%} & \multicolumn{4}{c}{48\%} & \multicolumn{4}{c}{52\%} \\
 & 300 & \multicolumn{4}{c}{62\%} & \multicolumn{4}{c}{57\%} & \multicolumn{4}{c}{61\%} \\
 & 500 & \multicolumn{4}{c}{76\%} & \multicolumn{4}{c}{65\%} & \multicolumn{4}{c}{75\%} \\ \hline
 &  & \multicolumn{2}{c}{$\phi^{+}_1=0.7$} & \multicolumn{2}{c|}{$\phi^{+}_2=0.2$} & \multicolumn{2}{c}{$\phi^{+}_1=0.7$} & \multicolumn{2}{c|}{$\phi^{*}_1=0.2$} & \multicolumn{2}{c}{$\phi^{*}_1=0.7$} & \multicolumn{2}{c}{$\phi^{*}_2=0.2$} \\ \hline
 & T & Mean & Std & Mean & \multicolumn{1}{c|}{Std} & Mean & Std & Mean & \multicolumn{1}{c|}{Std} & Mean & Std & Mean & Std \\ \hline
\multirow{3}{*}{$\nu=2$} & 100 & 0.711 & 0.133 & 0.174 & \multicolumn{1}{c|}{0.156} & 0.646 & 0.174 & 0.191 & \multicolumn{1}{c|}{0.180} & 0.709 & 0.123 & 0.184 & 0.140 \\
 & 300 & 0.690 & 0.069 & 0.206 & \multicolumn{1}{c|}{0.079} & 0.683 & 0.103 & 0.211 & \multicolumn{1}{c|}{0.114} & 0.700 & 0.085 & 0.191 & 0.093 \\
 & 500 & 0.707 & 0.050 & 0.206 & \multicolumn{1}{c|}{0.062} & 0.693 & 0.051 & 0.194 & \multicolumn{1}{c|}{0.067} & 0.696 & 0.063 & 0.196 & 0.068 \\
\multirow{3}{*}{$\nu=5$} & 100 & 0.732 & 0.146 & 0.168 & \multicolumn{1}{c|}{0.179} & 0.613 & 0.167 & 0.222 & \multicolumn{1}{c|}{0.217} & 0.746 & 0.131 & 0.133 & 0.161 \\
 & 300 & 0.719 & 0.085 & 0.186 & \multicolumn{1}{c|}{0.130} & 0.673 & 0.101 & 0.207 & \multicolumn{1}{c|}{0.132} & 0.679 & 0.115 & 0.220 & 0.132 \\
 & 500 & 0.711 & 0.078 & 0.198 & \multicolumn{1}{c|}{0.084} & 0.687 & 0.084 & 0.202 & \multicolumn{1}{c|}{0.122} & 0.703 & 0.090 & 0.191 & 0.096 \\ \hline
\end{tabular}
\end{table}

\subsection{The MMA($r',s'$) DGP}
We first consider the invertible MA(2,0) with parameters $\theta^{+}_1=0.7$ and $\theta^{+}_2=0.2$. Then we estimate the model assuming invertibility, noninvertibility MA(0,2), and mixed MMA(1,1). The identification rate corresponds to the number of times the estimation resulted in the invertible model MA(2,0). The mean and standard deviation correspond to the estimated parameters of the MA(2,0) model in the $M$ iterations. We next consider as DGP the noninvertible  MA(0,2) with parameters $\theta^{*}_1=0.7$ and $\theta^{*}_2=0.2$, and finally the MAR(1,1) with parameters $\theta^{+}_1=0.7$ and $\theta^{*}_1=0.2$. The same comparisons are made to establish identification rates and statistics on estimated parameters. The results are presented in Table \ref{ma_mc}.

\begin{table}[H]
\centering
\tiny
\caption{Identification rate, the mean and standard deviation of estimated parameters of MA(2,0), MMA(1,1), and MA(0,2) with errors alpha-stable and skew-t}
\vskip 0.2cm
\label{ma_mc}
\begin{tabular}{cccccccccccccc}
\hline
\multicolumn{1}{l}{} & \multicolumn{1}{l}{} & \multicolumn{12}{c}{\textbf{Panel A: Alpha-stable}} \\ \hline
\multicolumn{1}{l}{} & \multicolumn{1}{l}{} & \multicolumn{4}{c}{MA(2,0)} & \multicolumn{4}{c}{MMA(1,1)} & \multicolumn{4}{c}{MA(0,2)} \\ \cline{3-14} 
\multicolumn{1}{l}{Dist} & T & \multicolumn{4}{c}{Identification rate} & \multicolumn{4}{c}{Identification rate} & \multicolumn{4}{c}{Identification rate} \\ \hline
\multirow{3}{*}{$\alpha=1.2$} & 100 & \multicolumn{4}{c}{96\%} & \multicolumn{4}{c}{74\%} & \multicolumn{4}{c}{92\%} \\
 & 300 & \multicolumn{4}{c}{98\%} & \multicolumn{4}{c}{81\%} & \multicolumn{4}{c}{96\%} \\
 & 500 & \multicolumn{4}{c}{99\%} & \multicolumn{4}{c}{86\%} & \multicolumn{4}{c}{98\%} \\
\multirow{3}{*}{$\alpha=1.8$} & 100 & \multicolumn{4}{c}{66\%} & \multicolumn{4}{c}{52\%} & \multicolumn{4}{c}{64\%} \\
 & 300 & \multicolumn{4}{c}{80\%} & \multicolumn{4}{c}{69\%} & \multicolumn{4}{c}{82\%} \\
 & 500 & \multicolumn{4}{c}{87\%} & \multicolumn{4}{c}{76\%} & \multicolumn{4}{c}{88\%} \\ \hline
\multicolumn{1}{l}{} & \multicolumn{1}{l}{} & \multicolumn{2}{c}{$\theta^{+}_1=0.7$} & \multicolumn{2}{c|}{$\theta^{+}_2=0.2$} & \multicolumn{2}{c}{$\theta^{+}_1=0.7$} & \multicolumn{2}{c|}{$\theta^{*}_1=0.2$} & \multicolumn{2}{c}{$\theta^{*}_1=0.7$} & \multicolumn{2}{c}{$\theta^{*}_2=0.2$} \\ \cline{3-14} 
\multicolumn{1}{l}{Dist} & T & Mean & Std & Mean & \multicolumn{1}{c|}{Std} & Mean & Std & Mean & \multicolumn{1}{c|}{Std} & Mean & Std & Mean & Std \\ \hline
\multirow{3}{*}{$\alpha=1.2$} & 100 & 0.709 & 0.081 & 0.197 & \multicolumn{1}{c|}{0.128} & 0.674 & 0.251 & 0.168 & \multicolumn{1}{c|}{0.160} & 0.650 & 0.199 & 0.178 & 0.133 \\
 & \multicolumn{1}{l}{300} & 0.705 & 0.053 & 0.209 & \multicolumn{1}{c|}{0.059} & 0.728 & 0.108 & 0.171 & \multicolumn{1}{c|}{0.095} & 0.686 & 0.099 & 0.200 & 0.074 \\
 & \multicolumn{1}{l}{500} & 0.701 & 0.035 & 0.199 & \multicolumn{1}{c|}{0.041} & 0.709 & 0.087 & 0.201 & \multicolumn{1}{c|}{0.065} & 0.693 & 0.052 & 0.202 & 0.048 \\
\multirow{3}{*}{$\alpha=1.8$} & 100 & 0.725 & 0.137 & 0.217 & \multicolumn{1}{c|}{0.116} & 0.698 & 0.111 & 0.192 & \multicolumn{1}{c|}{0.156} & 0.670 & 0.115 & 0.188 & 0.122 \\
 & 300 & 0.695 & 0.111 & 0.195 & \multicolumn{1}{c|}{0.114} & 0.716 & 0.107 & 0.174 & \multicolumn{1}{c|}{0.127} & 0.707 & 0.073 & 0.215 & 0.074 \\
 & 500 & 0.710 & 0.076 & 0.204 & \multicolumn{1}{c|}{0.071} & 0.706 & 0.095 & 0.209 & \multicolumn{1}{c|}{0.077} & 0.703 & 0.047 & 0.205 & 0.063 \\ \hline
\multicolumn{1}{l}{} & \multicolumn{1}{l}{} & \multicolumn{1}{l}{} & \multicolumn{1}{l}{} & \multicolumn{1}{l}{} & \multicolumn{1}{l}{} & \multicolumn{1}{l}{} & \multicolumn{1}{l}{} & \multicolumn{1}{l}{} & \multicolumn{1}{l}{} & \multicolumn{1}{l}{} & \multicolumn{1}{l}{} & \multicolumn{1}{l}{} & \multicolumn{1}{l}{} \\ \hline
\multicolumn{1}{l}{} & \multicolumn{1}{l}{} & \multicolumn{12}{c}{\textbf{Panel B: Skew t}} \\ \hline
\multicolumn{1}{l}{} & \multicolumn{1}{l}{} & \multicolumn{4}{c}{MA(2,0)} & \multicolumn{4}{c}{MMA(1,1)} & \multicolumn{4}{c}{MA(0,2)} \\ \cline{3-14} 
\multicolumn{1}{l}{Dist} & T & \multicolumn{4}{c}{Identification rate} & \multicolumn{4}{c}{Identification rate} & \multicolumn{4}{c}{Identification rate} \\ \hline
\multirow{3}{*}{$\nu=2$} & 100 & \multicolumn{4}{c}{88\%} & \multicolumn{4}{c}{76\%} & \multicolumn{4}{c}{80\%} \\
 & 300 & \multicolumn{4}{c}{91\%} & \multicolumn{4}{c}{80\%} & \multicolumn{4}{c}{95\%} \\
 & 500 & \multicolumn{4}{c}{93\%} & \multicolumn{4}{c}{87\%} & \multicolumn{4}{c}{98\%} \\
\multirow{3}{*}{$\nu=5$} & 100 & \multicolumn{4}{c}{58\%} & \multicolumn{4}{c}{59\%} & \multicolumn{4}{c}{52\%} \\
 & 300 & \multicolumn{4}{c}{68\%} & \multicolumn{4}{c}{65\%} & \multicolumn{4}{c}{68\%} \\
 & 500 & \multicolumn{4}{c}{71\%} & \multicolumn{4}{c}{76\%} & \multicolumn{4}{c}{82\%} \\ \hline
 &  & \multicolumn{2}{c}{$\theta^{+}_1=0.7$} & \multicolumn{2}{c|}{$\theta^{+}_2=0.2$} & \multicolumn{2}{c}{$\theta^{+}_1=0.7$} & \multicolumn{2}{c|}{$\theta^{*}_1=0.2$} & \multicolumn{2}{c}{$\theta^{*}_1=0.7$} & \multicolumn{2}{c}{$\theta^{*}_2=0.2$} \\ \hline
 & T & Mean & Std & Mean & \multicolumn{1}{c|}{Std} & Mean & Std & Mean & \multicolumn{1}{c|}{Std} & Mean & Std & Mean & Std \\ \hline
\multirow{3}{*}{$\nu=2$} & 100 & 0.686 & 0.108 & 0.186 & \multicolumn{1}{c|}{0.103} & 0.682 & 0.105 & 0.187 & \multicolumn{1}{c|}{0.139} & 0.669 & 0.092 & 0.203 & 0.128 \\
 & 300 & 0.694 & 0.080 & 0.194 & \multicolumn{1}{c|}{0.084} & 0.722 & 0.104 & 0.188 & \multicolumn{1}{c|}{0.084} & 0.680 & 0.081 & 0.187 & 0.079 \\
 & 500 & 0.697 & 0.076 & 0.202 & \multicolumn{1}{c|}{0.067} & 0.702 & 0.091 & 0.194 & \multicolumn{1}{c|}{0.790} & 0.686 & 0.063 & 0.202 & 0.059 \\
\multirow{3}{*}{$\nu=5$} & 100 & 0.683 & 0.109 & 0.204 & \multicolumn{1}{c|}{0.122} & 0.678 & 0.159 & 0.215 & \multicolumn{1}{c|}{0.161} & 0.656 & 0.112 & 0.159 & 0.110 \\
 & 300 & 0.686 & 0.062 & 0.194 & \multicolumn{1}{c|}{0.092} & 0.679 & 0.117 & 0.213 & \multicolumn{1}{c|}{0.098} & 0.688 & 0.068 & 0.188 & 0.070 \\
 & 500 & 0.697 & 0.060 & 0.201 & \multicolumn{1}{c|}{0.061} & 0.689 & 0.074 & 0.205 & \multicolumn{1}{c|}{0.071} & 0.692 & 0.056 & 0.203 & 0.057 \\ \hline
\end{tabular}
\end{table}

\subsection{The MARMA($r,s,r',s'$) DGP}
For the MARMA, the DGP is first the causal and invertible model MARMA(1,0,1,0), that is $p=1$ and $q=1$, with parameters $\phi^{+}_1=0.7$ and $\theta^{*}_1=0.2$. We then estimate the model assuming causality and invertibility MARMA($1,0,1,0$), noncausality and invertibility MARMA(0,1,1,0), causality and noninvertibility MARMA($1,0,0,1$) and finally noncausality and noninvertibility MARMA($0,1,0,1$). Subsequently, we consider as DGP the noncausal and invertible model MARMA($0,1,1,0$) with parameters $\phi^{*}_1=0.7$ and $\theta^{+}_1=0.2$, then the causal and noninvertible model MARMA(1,0,0,1) with parameters $\phi^{+}_1=0.7$ and $\theta^{*}_1=0.2$, and finally the noncausal and noninvertible model MARMA(0,1,0,1) with parameters $\phi^{*}_1=0.7$ and $\theta^{*}_1=0.2$. The same comparisons are made to establish the identification rate in all cases. In each case, we report the mean and standard deviation of the estimated parameters.  The results are presented in Table \ref{marma_mc}.

\begin{table}[h]
\centering
\tiny
\caption{Identification rate, the mean and standard deviation of estimated parameters of the MARMA(1,0,1,0), MARMA(0,1,1,0), MARMA(1,0,0,1), and MARMA(0,1,0,1) with errors alpha-stable and skew-t}
\label{marma_mc}
\resizebox{\textwidth}{!}{%
\begin{tabular}{cccccccccccccccccc}
\hline
\multicolumn{1}{l}{} & \multicolumn{1}{l}{} & \multicolumn{16}{c}{\textbf{Panel A: Alpha-stable}} \\ \hline
\multicolumn{1}{l}{} & \multicolumn{1}{l}{} & \multicolumn{4}{c}{MARMA(1,0,1,0)} & \multicolumn{4}{c}{MARMA(0,1,1,0)} & \multicolumn{4}{c}{MARMA(1,0,0,1)} & \multicolumn{4}{c}{MARMA(0,1,0,1)} \\ \cline{3-18} 
\multicolumn{1}{l}{Dist} & T & \multicolumn{4}{c}{Identification rate} & \multicolumn{4}{c}{Identification rate} & \multicolumn{4}{c}{Identification rate} & \multicolumn{4}{c}{Identification rate} \\ \hline
\multirow{3}{*}{$\alpha=1.2$} & 100 & \multicolumn{4}{c}{78\%} & \multicolumn{4}{c}{80\%} & \multicolumn{4}{c}{84\%} & \multicolumn{4}{c}{88\%} \\
 & 300 & \multicolumn{4}{c}{92\%} & \multicolumn{4}{c}{98\%} & \multicolumn{4}{c}{98\%} & \multicolumn{4}{c}{92\%} \\
 & 500 & \multicolumn{4}{c}{96\%} & \multicolumn{4}{c}{99\%} & \multicolumn{4}{c}{98\%} & \multicolumn{4}{c}{95\%} \\
\multirow{3}{*}{$\alpha=1.8$} & 100 & \multicolumn{4}{c}{54\%} & \multicolumn{4}{c}{47\%} & \multicolumn{4}{c}{48\%} & \multicolumn{4}{c}{51\%} \\
 & 300 & \multicolumn{4}{c}{61\%} & \multicolumn{4}{c}{70\%} & \multicolumn{4}{c}{70\%} & \multicolumn{4}{c}{63\%} \\
 & 500 & \multicolumn{4}{c}{74\%} & \multicolumn{4}{c}{82\%} & \multicolumn{4}{c}{78\%} & \multicolumn{4}{c}{75\%} \\ \hline
\multicolumn{1}{l}{} & \multicolumn{1}{l}{} & \multicolumn{2}{c}{$\phi^{+}_1=0.7$} & \multicolumn{2}{c|}{$\theta^{+}_1=0.2$} & \multicolumn{2}{c}{$\phi^{*}_1=0.7$} & \multicolumn{2}{c|}{$\theta^{+}_1=0.2$} & \multicolumn{2}{c}{$\phi^{+}_1=0.7$} & \multicolumn{2}{l|}{$\theta^{*}_1=0.2$} & \multicolumn{2}{l}{$\phi^{*}_1=0.7$} & \multicolumn{2}{l}{$\theta^{*}_1=0.2$} \\ \cline{3-18} 
\multicolumn{1}{l}{Dist} & T & Mean & Std & Mean & \multicolumn{1}{c|}{Std} & Mean & Std & Mean & \multicolumn{1}{c|}{Std} & Mean & \multicolumn{1}{l}{std} & Mean & \multicolumn{1}{l|}{std} & Mean & \multicolumn{1}{l}{std} & Mean & \multicolumn{1}{l}{std} \\ \hline
\multirow{3}{*}{$\alpha=1.2$} & 100 & 0.666 & 0.118 & 0.209 & \multicolumn{1}{c|}{0.150} & 0.703 & 0.061 & 0.189 & \multicolumn{1}{c|}{0.099} & 0.682 & 0.095 & 0.205 & \multicolumn{1}{c|}{0.096} & 0.674 & 0.076 & 0.225 & 0.106 \\
 & \multicolumn{1}{l}{300} & 0.708 & 0.040 & 0.212 & \multicolumn{1}{c|}{0.104} & 0.699 & 0.041 & 0.203 & \multicolumn{1}{c|}{0.049} & 0.701 & 0.042 & 0.210 & \multicolumn{1}{c|}{0.069} & 0.696 & 0.043 & 0.187 & 0.099 \\
 & \multicolumn{1}{l}{500} & 0.700 & 0.035 & 0.196 & \multicolumn{1}{c|}{0.094} & 0.701 & 0.032 & 0.200 & \multicolumn{1}{c|}{0.047} & 0.700 & 0.038 & 0.201 & \multicolumn{1}{c|}{0.045} & 0.704 & 0.039 & 0.197 & 0.057 \\
\multirow{3}{*}{$\alpha=1.8$} & 100 & 0.680 & 0.078 & 0.243 & \multicolumn{1}{c|}{0.162} & 0.667 & 0.109 & 0.223 & \multicolumn{1}{c|}{0.170} & 0.709 & 0.095 & 0.169 & \multicolumn{1}{c|}{0.182} & 0.706 & 0.087 & 0.211 & 0.185 \\
 & 300 & 0.691 & 0.065 & 0.214 & \multicolumn{1}{c|}{0.131} & 0.689 & 0.053 & 0.206 & \multicolumn{1}{c|}{0.085} & 0.697 & 0.054 & 0.217 & \multicolumn{1}{c|}{0.100} & 0.693 & 0.052 & 0.196 & 0.100 \\
 & 500 & 0.702 & 0.039 & 0.198 & \multicolumn{1}{c|}{0.073} & 0.693 & 0.050 & 0.203 & \multicolumn{1}{c|}{0.080} & 0.705 & 0.041 & 0.208 & \multicolumn{1}{c|}{0.076} & 0.708 & 0.039 & 0.205 & 0.081 \\ \hline
\multicolumn{1}{l}{} & \multicolumn{1}{l}{} & \multicolumn{1}{l}{} & \multicolumn{1}{l}{} & \multicolumn{1}{l}{} & \multicolumn{1}{l}{} & \multicolumn{1}{l}{} & \multicolumn{1}{l}{} & \multicolumn{1}{l}{} & \multicolumn{1}{l}{} & \multicolumn{1}{l}{} & \multicolumn{1}{l}{} & \multicolumn{1}{l}{} & \multicolumn{1}{l}{} & \multicolumn{1}{l}{} & \multicolumn{1}{l}{} & \multicolumn{1}{l}{} & \multicolumn{1}{l}{} \\ \hline
\multicolumn{1}{l}{} & \multicolumn{1}{l}{} & \multicolumn{16}{c}{\textbf{Panel B: Skew t}} \\ \hline
\multicolumn{1}{l}{} & \multicolumn{1}{l}{} & \multicolumn{4}{c}{MARMA(1,0,1,0)} & \multicolumn{4}{c}{MARMA(0,1,1,0)} & \multicolumn{4}{c}{MARMA(1,0,0,1)} & \multicolumn{4}{c}{MARMA(0,1,0,1)} \\ \cline{3-18} 
\multicolumn{1}{l}{Dist} & T & \multicolumn{4}{c}{Identification rate} & \multicolumn{4}{c}{Identification rate} & \multicolumn{4}{c}{Identification rate} & \multicolumn{4}{c}{Identification rate} \\ \hline
\multirow{3}{*}{$\nu=2$} & 100 & \multicolumn{4}{c}{65\%} & \multicolumn{4}{c}{64\%} & \multicolumn{4}{c}{69\%} & \multicolumn{4}{c}{64\%} \\
 & 300 & \multicolumn{4}{c}{84\%} & \multicolumn{4}{c}{81\%} & \multicolumn{4}{c}{86\%} & \multicolumn{4}{c}{80\%} \\
 & 500 & \multicolumn{4}{c}{90\%} & \multicolumn{4}{c}{86\%} & \multicolumn{4}{c}{96\%} & \multicolumn{4}{c}{88\%} \\
\multirow{3}{*}{$\nu=5$} & 100 & \multicolumn{4}{c}{42\%} & \multicolumn{4}{c}{56\%} & \multicolumn{4}{c}{66\%} & \multicolumn{4}{c}{51\%} \\
 & 300 & \multicolumn{4}{c}{58\%} & \multicolumn{4}{c}{61\%} & \multicolumn{4}{c}{73\%} & \multicolumn{4}{c}{58\%} \\
 & 500 & \multicolumn{4}{c}{71\%} & \multicolumn{4}{c}{68\%} & \multicolumn{4}{c}{88\%} & \multicolumn{4}{c}{70\%} \\ \hline
 &  & \multicolumn{2}{c}{$\phi^{+}_1=0.7$} & \multicolumn{2}{c|}{$\theta^{+}_1=0.2$} & \multicolumn{2}{c}{$\phi^{*}_1=0.7$} & \multicolumn{2}{c|}{$\theta^{+}_1=0.2$} & \multicolumn{2}{c}{$\phi^{+}_1=0.7$} & \multicolumn{2}{l|}{$\theta^{*}_1=0.2$} & \multicolumn{2}{l}{$\phi^{*}_1=0.7$} & \multicolumn{2}{l}{$\theta^{*}_1=0.2$} \\ \hline
 & T & Mean & Std & Mean & \multicolumn{1}{c|}{Std} & Mean & Std & Mean & \multicolumn{1}{c|}{Std} & Mean & Std & Mean & \multicolumn{1}{c|}{Std} & Mean & Std & Mean & Std \\ \hline
\multirow{3}{*}{$\nu=2$} & 100 & 0.708 & 0.086 & 0.187 & \multicolumn{1}{c|}{0.143} & 0.693 & 0.096 & 0.185 & \multicolumn{1}{c|}{0.141} & 0.673 & 0.128 & 0.205 & \multicolumn{1}{c|}{0.167} & 0.669 & 0.145 & 0.212 & 0.179 \\
 & 300 & 0.699 & 0.050 & 0.202 & \multicolumn{1}{c|}{0.097} & 0.698 & 0.082 & 0.211 & \multicolumn{1}{c|}{0.086} & 0.698 & 0.057 & 0.200 & \multicolumn{1}{c|}{0.097} & 0.682 & 0.061 & 0.209 & 0.103 \\
 & 500 & 0.696 & 0.046 & 0.203 & \multicolumn{1}{c|}{0.086} & 0.700 & 0.043 & 0.193 & \multicolumn{1}{c|}{0.076} & 0.702 & 0.047 & 0.201 & \multicolumn{1}{c|}{0.077} & 0.704 & 0.053 & 0.201 & 0.086 \\
\multirow{3}{*}{$\nu=5$} & 100 & 0.692 & 0.091 & 0.193 & \multicolumn{1}{c|}{0.168} & 0.712 & 0.128 & 0.159 & \multicolumn{1}{c|}{0.209} & 0.682 & 0.132 & 0.207 & \multicolumn{1}{c|}{0.170} & 0.662 & 0.156 & 0.212 & 0.185 \\
 & 300 & 0.694 & 0.059 & 0.198 & \multicolumn{1}{c|}{0.093} & 0.701 & 0.063 & 0.206 & \multicolumn{1}{c|}{0.095} & 0.699 & 0.055 & 0.198 & \multicolumn{1}{c|}{0.128} & 0.696 & 0.071 & 0.205 & 0.099 \\
 & 500 & 0.699 & 0.052 & 0.201 & \multicolumn{1}{c|}{0.079} & 0.700 & 0.054 & 0.205 & \multicolumn{1}{c|}{0.079} & 0.702 & 0.036 & 0.197 & \multicolumn{1}{c|}{0.080} & 0.699 & 0.063 & 0.199 & 0.085 \\ \hline
\end{tabular}%
}
\end{table}

\subsection{Monte Carlo outcomes}

In Tables \ref{ar_mc}, \ref{ma_mc}, it is obvious that the model identification is a function of the level at which the data depart from normality. For example, for the alpha-stable distribution with $\alpha=1.2$ and the skew-t with $\nu=2$, the identification rate is significantly higher than in the case of $\alpha=1.8$ and $\nu=5$, being the latter close to the normal distribution, with thinner tails. In all cases, the identification rate increases as $T$ increases. Note that when the $T=100$ for $\alpha=1.8$ and $\nu=5$, the identification is close to 50\%. This indicates that the model cannot be accurately identified. The empirical bias of the estimated parameters and the standard deviation also depends on the leptokurtocity of the generated data. As it departs from normality, there is less bias and less standard deviation. In addition, the empirical bias and standard deviation decrease as $T$ increases. For the AR and MA components, the 0.2 parameter has a larger bias than the 0.7 parameter. Table \ref{marma_mc} exhibits a correct estimation and identification of the MARMA model, both for the AR and the MA components, regardless of the location of the roots, outside or inside the unit circle. For the four models, the identification is similar.

\section{Empirical application}
\subsection{Data description}
We consider monthly returns of 24 Famma-French emerging market stocks portfolios observed from January 1990 to August 2022, for a total of 392 months. The data set is taken from Kenneth French's website \url{http://mba.tuck.dartmouth.edu/pages/faculty/ken.french/data_library.html}. The emerging markets countries currently included are Argentina, Brazil, Chile, China, Colombia, Czech Republic, Egypt, Greece, Hungary, India, Indonesia, Malaysia, Mexico, Pakistan, Peru, Philippines, Poland, Qatar, Russia, Saudi Arabia, South Africa, South Korea, Taiwan, Thailand, Turkey, United Arab Emirates.

\begin{center}
     \begin{figure}[h]
         \centering         
           \caption{Size and B/M portfolios}
           \vskip 0.2cm
         \includegraphics[width=0.8 \textwidth]{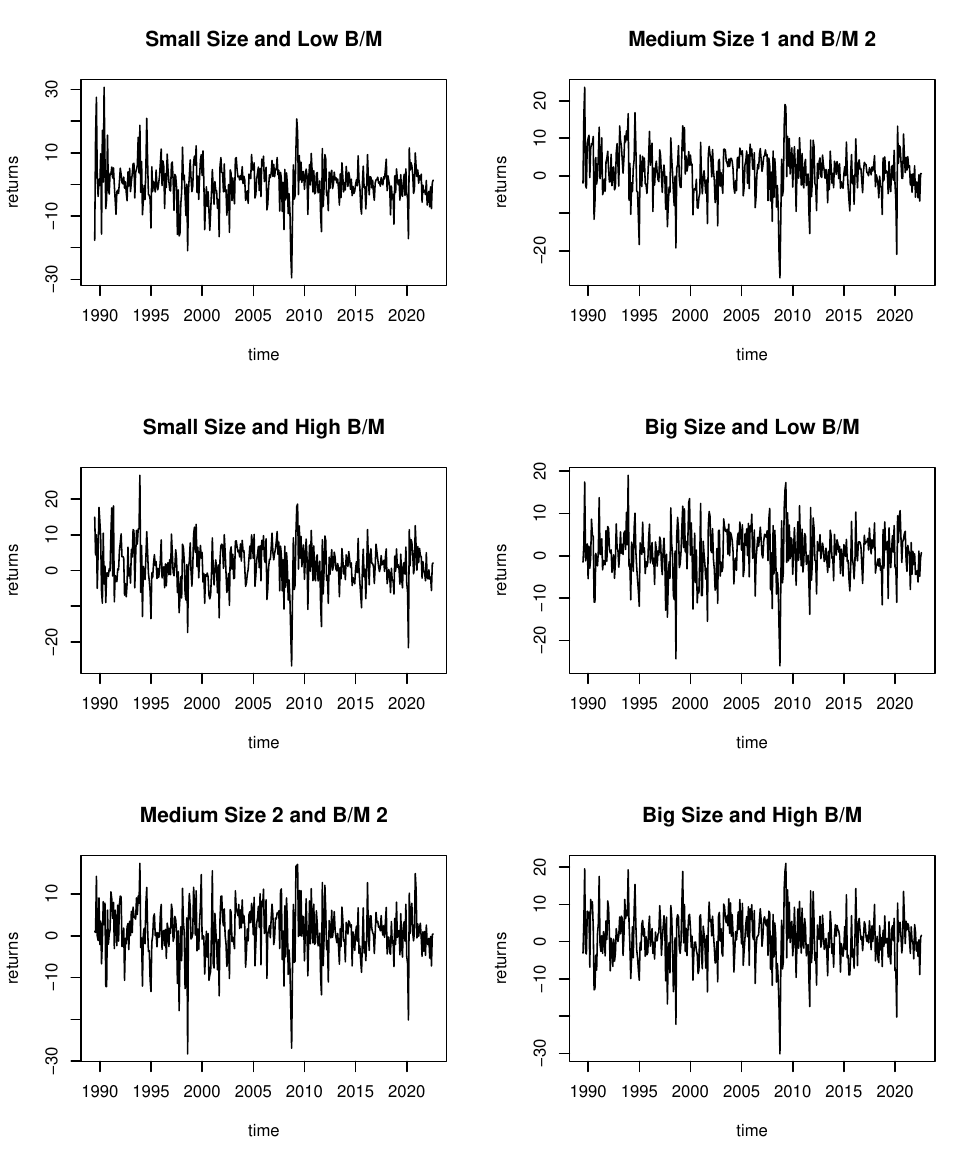}
       
        \label{ff_1}
     \end{figure}
 \end{center}

\begin{center}
     \begin{figure}[h]
         \centering         
           \caption{Size and Operating Profit portfolios}
           \vskip 0.2cm
     \includegraphics[width=0.8 \textwidth]{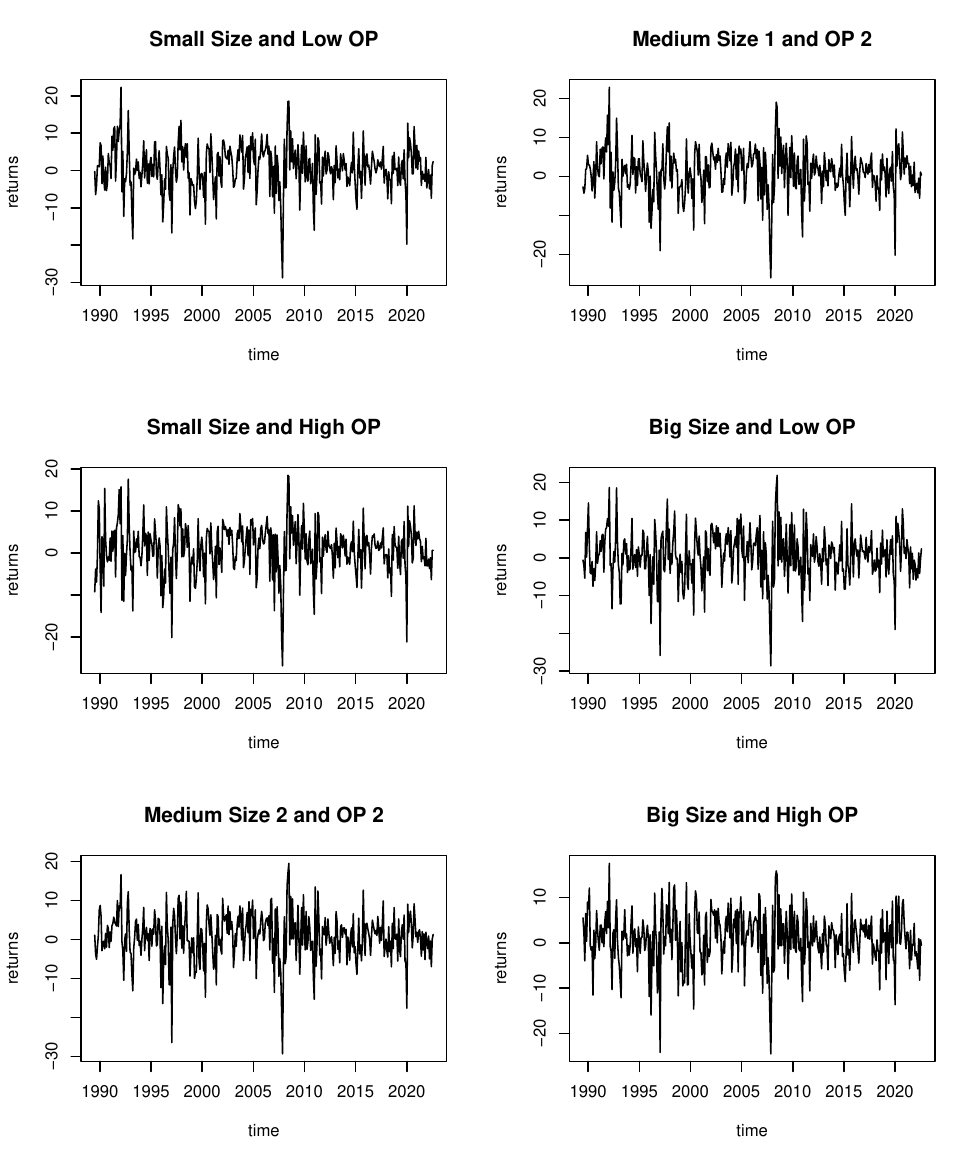}
       
        \label{ff_2}
     \end{figure}
 \end{center}

\begin{center}
     \begin{figure}[h]
         \centering         
           \caption{Size and Investment portfolios}
           \vskip 0.2cm
     \includegraphics[width=0.8 \textwidth]{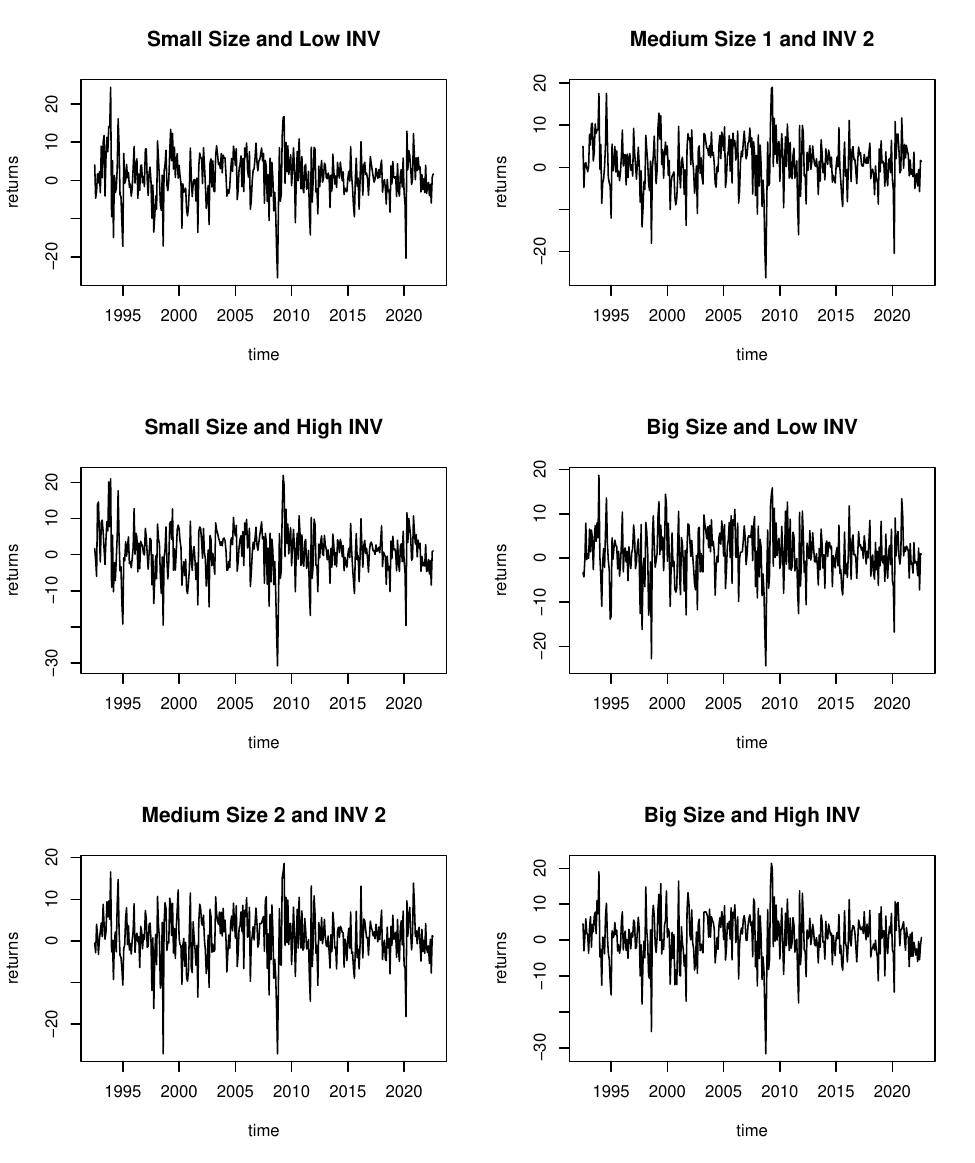}
        \label{ff_3}
     \end{figure}
 \end{center}

\begin{center}
     \begin{figure}[h]
         \centering         
           \caption{Size and Momentum portfolios}
           \vskip 0.2cm
     \includegraphics[width=0.8 \textwidth]{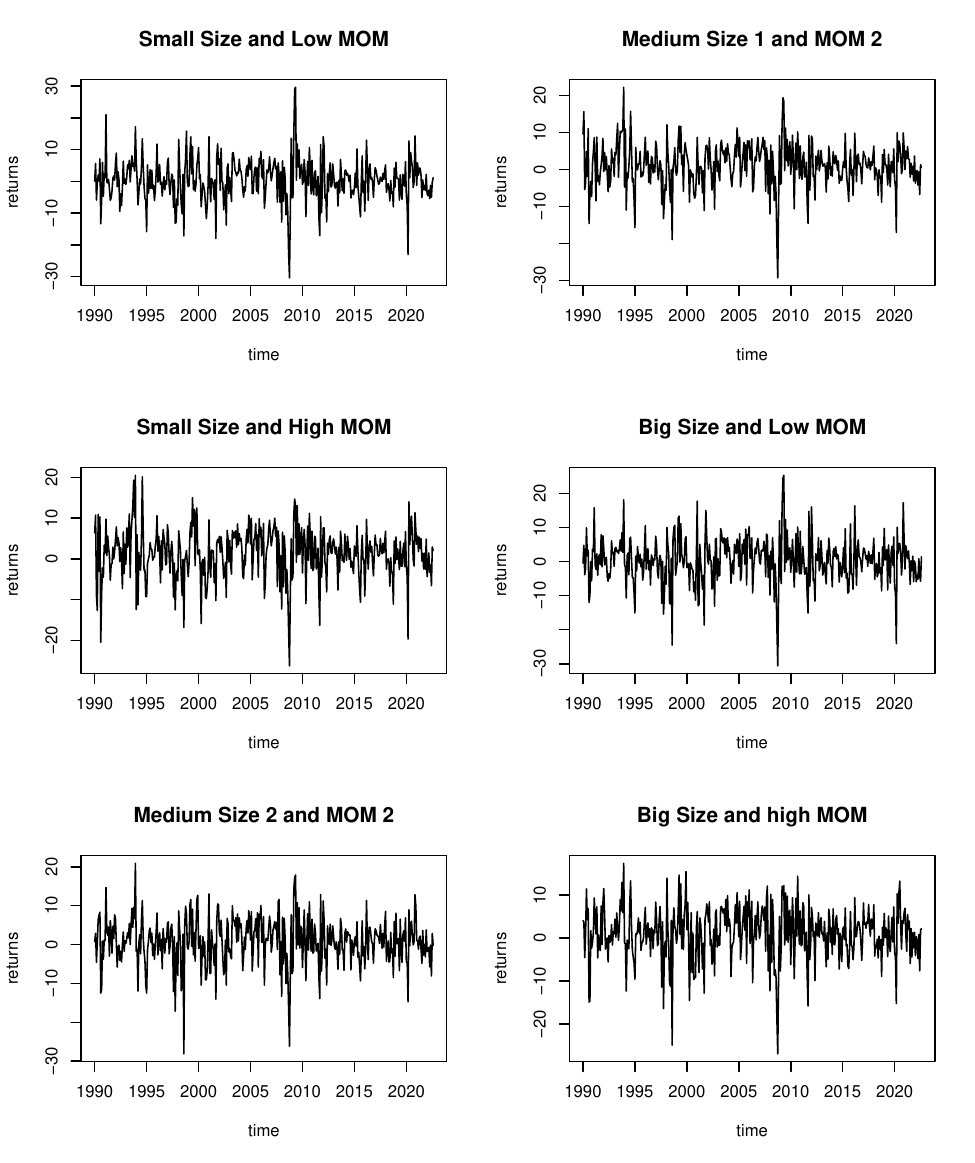}
        \label{ff_4}
     \end{figure}
 \end{center}

All returns are in U.S. dollars, include dividends and capital gains, and are not continuously compounded. We consider several categories of returns. In the first category, the stocks are sorted by two Market-cap (size) and three book-to-market equity (B/M) groups, leading to six average value-weighted portfolios. The returns of The returns of the portfolios are in Figure \ref{ff_1}. Big-size stocks are those in the country's top 90\% of the market cap, and small stocks are those in the bottom 10\%. The B/M breakpoints for big and small stocks in a country are the 30th and 70th percentiles of B/M for the big stocks of the country. In the second category, the stocks are sorted by two sizes and three profitability (OP) groups, leading to six average value-weighted portfolios. The returns of the portfolios are in Figure \ref{ff_2}. The OP breakpoints for big and small stocks in a country are the 30th and 70th percentiles of OP for the big stocks of the country. In the third category, the stocks are sorted by two sizes and three Investment (Inv) groups, leading to six average value-weighted portfolios, The returns of the portfolios are in Figure \ref{ff_3}. The Inv breakpoints for big and small stocks in a country are the 30th and 70th percentiles of Inv for the big stocks of the country. In the fourth category, the stocks are sorted by two market sizes and three lagged momentum return groups (MOM), leading to six average value-weighted portfolios, The returns of the portfolios are in Figure \ref{ff_4}. For portfolios formed at the end of the month $t–1$, the lagged momentum return is a stock's cumulative return for $t–12$ to $t–2$. The momentum breakpoints for all stocks in a country are the 30th and 70th percentiles of the lagged momentum returns of the country's big stocks.

We selected the portfolios in emerging markets since the dynamics are stronger than those found in developed markets, as they are exposed to a greater extent to risk factors such as interest rate variations, currency devaluation, and policy changes between governments, among others. For an application in developed markets, see \cite{gospodinov2015minimum} where all portfolios exhibit noninvertible MA(0,1) dynamics.

\subsection{Data features}

Although the constructed portfolios are formed from assets in competitive markets and are diversified among different countries and sectors, they exhibit characteristics of time dependency that MARMA models can capture. In addition, the portfolios exhibit non-Gaussian characteristics such as skewness and kurtosis, which we can observe in Table \ref{desc}.  
 
We perform a preliminary estimation of ARMA models by Gaussian-likelihood to select lag orders, assuming consequently causality and invertibility. To identify the dynamics, we rely on the BIC, the significance of the lags in the ACF and the PACF, as well as the significance of the coefficients in each possible model. We restrict the order of the AR component to $p=2$ and the MA component to $q=2$. In Table \ref{desc}, we report the mean, the standard deviation, the skewness, the kurtosis, and the identification of the ARMA model using Gaussian-likelihood estimation.

\begin{table}[H]
\centering
\caption{Descriptive Statistics and Dynamics}
\vskip 0.2cm
\label{desc}
\tiny
\begin{tabular}{lccccc}
\cline{2-6}
 & \multicolumn{5}{c}{Size and Book-to-Market Portfolios} \\ \cline{2-6} 
 & Mean & Sd.Dev. & Skewness & Kurtosis & Dynamics \\ \hline
Small   Size and Low B/M & 0.005 & 0.067 & -0.107 & 6.095 & AR(2) \\
Medium Size 1 and B/M 2 & 0.010 & 0.062 & -0.408 & 5.017 & AR(2) \\
Small size and High B/M & 0.014 & 0.063 & -0.233 & 4.846 & AR(2) \\
Big Size and Low B/M & 0.007 & 0.059 & -0.461 & 4.795 & AR(2) \\
Medium Size 2 and B/M 2 & 0.009 & 0.063 & -0.584 & 4.955 & AR(2) \\
Big Size and High B/M & 0.010 & 0.066 & -0.366 & 4.633 & AR(1) \\ \hline
 & \multicolumn{5}{c}{Size and Operating Profitability Portfolios} \\ \cline{2-6} 
 & Mean & Sd.Dev. & Skewness & Kurtosis & Dynamics \\ \hline
Small   Size and Low OP & 0.008 & 0.061 & -0.492 & 5.147 & AR(2) \\
Medium Size 1 and OP 2 & 0.011 & 0.059 & -0.488 & 5.142 & AR(2) \\
Small Size and High OP & 0.009 & 0.060 & -0.611 & 4.981 & AR(2) \\
Big Size and Low OP & 0.006 & 0.066 & -0.401 & 4.768 & AR(2) \\
Medium Size 2 and OP 2 & 0.007 & 0.061 & -0.668 & 5.513 & AR(2) \\
Big Size and High OP & 0.009 & 0.060 & -0.519 & 4.386 & AR(2) \\ \cline{2-6} 
 & \multicolumn{5}{c}{Size and Investment Portfolios} \\ \cline{2-6} 
 & Mean & Sd.Dev. & Skewness & Kurtosis & \multicolumn{1}{l}{Dynamics} \\ \hline
Small   Size and Low INV & 0.010 & 0.061 & -0.434 & 4.849 & AR(2) \\
Medium Size 1 and INV 2 & 0.011 & 0.059 & -0.537 & 5.125 & AR(2) \\
Small Size and High INV & 0.007 & 0.065 & -0.477 & 5.340 & AR(2) \\
Big Size and Low INV & 0.009 & 0.059 & -0.529 & 4.583 & AR(2) \\
Medium Size 2 and INV 2 & 0.008 & 0.061 & -0.592 & 5.195 & AR(2) \\
Big Size and High INV & 0.007 & 0.067 & -0.526 & 5.202 & AR(2) \\ \cline{2-6} 
 & \multicolumn{5}{c}{Size and Momentum Portfolios} \\ \cline{2-6} 
 & Mean & Sd.Dev. & Skewness & Kurtosis & Dynamics \\ \hline
Small Size and Low   MOM & 0.004 & 0.066 & -0.041 & 6.162 & AR(1) \\
Medium Size 1 and MOM 2 & 0.012 & 0.059 & -0.514 & 5.571 & AR(2) \\
Small Size and High MOM & 0.013 & 0.062 & -0.549 & 4.641 & AR(2) \\
Big Size and Low MOM & 0.004 & 0.067 & -0.256 & 5.481 & AR(2) \\
Medium Size 2 and MOM 2 & 0.008 & 0.061 & -0.542 & 5.138 & AR(1) \\
Big Size and high MOM & 0.011 & 0.061 & -0.687 & 4.793 & AR(2) \\ \hline
\end{tabular}
\end{table}

 Overall the mean of the returns is close to zero. The standard deviations are in the range of 0.059 and 0.067, in annual terms between 20\% and 23\%. All portfolios have negative skewness and excess kurtosis. The low B/M, low OP, and low MOM portfolios display higher kurtosis than the high B/M, high OP, and high MOM portfolios. On the contrary, the low INV portfolio has a lower kurtosis than the high INV portfolio. It is found that all returns follow autoregressive processes without a moving average component. Out of twenty-four portfolios, twenty-one are AR(2), and three are AR(1).  
 \vskip 0.15cm
\begin{table}[h]

\centering
\caption{Estimation of portfolios returns using $R_T(\boldsymbol{\vartheta})$}
\vskip 0.2cm
\label{est1}
\tiny
\begin{tabular}{lccccc}
\hline
 & \multicolumn{1}{l}{} & \multicolumn{2}{c}{\textbf{Causal}} & \multicolumn{2}{c}{\textbf{Noncausal}} \\ \hline
\textit{\textbf{Size and Book-to-Market Portfolios}} & Model & $\phi^{+}_1$ & $\phi^{+}_2$ & $\phi^{*}_1$ & $\phi^{*}_2$ \\ \hline
Small Size and Low B/M & MAR(1,1) & \begin{tabular}[c]{@{}c@{}}-0.368\\ (0.046)\end{tabular} &  & \begin{tabular}[c]{@{}c@{}}0.506\\ (0.049)\end{tabular} &  \\
Medium Size 1 and B/M 2 & MAR(1,1) & \begin{tabular}[c]{@{}c@{}}-0.154\\ (0.046)\end{tabular} &  & \begin{tabular}[c]{@{}c@{}}0.438\\ (0.051)\end{tabular} &  \\
Small Size and High B/M & MAR(1,1) & \begin{tabular}[c]{@{}c@{}}-0.191\\ (0.052)\end{tabular} &  & \begin{tabular}[c]{@{}c@{}}0.434\\ (0.047)\end{tabular} &  \\
Big Size and Low B/M & MAR(1,1) & \begin{tabular}[c]{@{}c@{}}-0.222\\ (0.051)\end{tabular} &  & \begin{tabular}[c]{@{}c@{}}0.429\\ (0.047)\end{tabular} &  \\
Medium Size 2 and B/M 2 & MAR(0,2) &  &  & \begin{tabular}[c]{@{}c@{}}0.168\\ (0.051)\end{tabular} & \begin{tabular}[c]{@{}c@{}}0.154\\ (0.051)\end{tabular} \\
Big Size and High B/M & MAR(0,1) &  &  & \begin{tabular}[c]{@{}c@{}}0.242\\ (0.051)\end{tabular} &  \\ \hline
 & \multicolumn{1}{l}{} & \multicolumn{2}{c}{\textbf{Causal}} & \multicolumn{2}{c}{\textbf{Noncausal}} \\ \hline
\textit{\textbf{Size and Operating Profitability Portfolios}} & Model & $\phi^{+}_1$ & $\phi^{+}_2$ & $\phi^{*}_1$ & $\phi^{*}_2$ \\ \hline
Small   Size and Low OP & MAR(1,1) & \begin{tabular}[c]{@{}c@{}}-0.125\\ (0.053)\end{tabular} &  & \begin{tabular}[c]{@{}c@{}}0.455\\ (0.048)\end{tabular} &  \\
Medium Size 1 and OP 2 & MAR(1,1) & \begin{tabular}[c]{@{}c@{}}-0.186\\ (0.053)\end{tabular} &  & \begin{tabular}[c]{@{}c@{}}0.461\\ (0.048)\end{tabular} &  \\
Small Size and High OP & MAR(1,1) & \begin{tabular}[c]{@{}c@{}}-0.155\\ (0.053)\end{tabular} &  & \begin{tabular}[c]{@{}c@{}}0.405\\ (0.049)\end{tabular} &  \\
Big Size and Low OP & MAR(0,2) &  &  & \begin{tabular}[c]{@{}c@{}}0.199\\ (0.053)\end{tabular} & \begin{tabular}[c]{@{}c@{}}0.120\\ (0.053)\end{tabular} \\
Medium Size 2 and OP 2 & MAR(0,2) &  &  & \begin{tabular}[c]{@{}c@{}}0.205\\ (0.052)\end{tabular} & \begin{tabular}[c]{@{}c@{}}0.163\\ (0.053)\end{tabular} \\
Big Size and High OP & MAR(0,2) &  &  & \begin{tabular}[c]{@{}c@{}}0.177\\ (0.053)\end{tabular} & \begin{tabular}[c]{@{}c@{}}0.141\\ (0.053)\end{tabular} \\ \hline
 & \multicolumn{1}{l}{} & \multicolumn{2}{c}{\textbf{Causal}} & \multicolumn{2}{c}{\textbf{Noncausal}} \\ \hline
\textit{\textbf{Size and Investment Portfolios}} & Model & $\phi^{+}_1$ & $\phi^{+}_2$ & $\phi^{*}_1$ & $\phi^{*}_2$ \\ \hline
Small   Size and Low INV & MAR(1,1) & \begin{tabular}[c]{@{}c@{}}-0.138\\ (0.053)\end{tabular} &  & \begin{tabular}[c]{@{}c@{}}0.417\\ (0.054)\end{tabular} &  \\
Medium Size 1 and INV 2 & MAR(1,1) & \begin{tabular}[c]{@{}c@{}}-0.185\\ (0.053)\end{tabular} &  & \begin{tabular}[c]{@{}c@{}}0.486\\ (0.048)\end{tabular} &  \\
Small Size and High INV & MAR(1,1) & \begin{tabular}[c]{@{}c@{}}-0.153\\ (0.054)\end{tabular} & \multicolumn{1}{l}{} & \begin{tabular}[c]{@{}c@{}}0.464\\ (0.048)\end{tabular} &  \\
Big Size and Low INV & MAR(0,2) &  &  & \begin{tabular}[c]{@{}c@{}}0.193\\ (0.053)\end{tabular} & \begin{tabular}[c]{@{}c@{}}0.111\\ (0.054)\end{tabular} \\
Medium Size 2 and INV 2 & MAR(0,2) &  &  & \begin{tabular}[c]{@{}c@{}}0.180\\ (0.053)\end{tabular} & \begin{tabular}[c]{@{}c@{}}0.160\\ (0.054)\end{tabular} \\
Big Size and High INV & MAR(0,2) &  &  & \begin{tabular}[c]{@{}c@{}}0.229\\ (0.053)\end{tabular} & \begin{tabular}[c]{@{}c@{}}0.132\\ (0.054)\end{tabular} \\ \hline
 & \multicolumn{1}{l}{} & \multicolumn{2}{c}{\textbf{Causal}} & \multicolumn{2}{c}{\textbf{Noncausal}} \\ \hline
\textit{\textbf{Size and Momentum Portfolios}} & Model & $\phi^{+}_1$ & $\phi^{+}_2$ & $\phi^{*}_1$ & $\phi^{*}_2$ \\ \hline
Small Size and Low MOM & MAR(0,1) &  &  & \begin{tabular}[c]{@{}c@{}}0.286\\ (0.051)\end{tabular} &  \\
Medium Size 1 and MOM 2 & MAR(0,2) &  &  & \begin{tabular}[c]{@{}c@{}}0.251\\ (0.051)\end{tabular} & \begin{tabular}[c]{@{}c@{}}0.110\\ (0.052)\end{tabular} \\
Small Size and High MOM & MAR(0,2) &  &  & \begin{tabular}[c]{@{}c@{}}0.229\\ (0.051)\end{tabular} & \begin{tabular}[c]{@{}c@{}}0.107\\ (0.052)\end{tabular} \\
Big Size and Low MOM & MAR(1,1) & \begin{tabular}[c]{@{}c@{}}-0.212\\ (0.051)\end{tabular} &  & \begin{tabular}[c]{@{}c@{}}0.406\\ (0.048)\end{tabular} &  \\
Medium Size 2 and MOM 2 & MAR(0,1) &  &  & \begin{tabular}[c]{@{}c@{}}0.213\\ (0.051)\end{tabular} &  \\
Big Size and high MOM & MAR(0,2) &  &  & \begin{tabular}[c]{@{}c@{}}0.187\\ (0.051)\end{tabular} & \begin{tabular}[c]{@{}c@{}}0.139\\ (0.052)\end{tabular} \\ \hline
\end{tabular}
\end{table}
\vskip 0.15cm

\subsection{MARMA identification}
Table \ref{est1} reports the estimation using $R_T(\boldsymbol{\vartheta})$. We present the identified models as well as the estimated parameters with the standard errors computed as in Appendix A.
The tests for i.i.d and the \textit{p-value} for the first five lags of the Ljung-Box test are reported in Table \ref{tests}.

We found noncausal features in all of the twenty-four portfolios. This result indicates that these portfolios have the characteristic of adjusting adequately to near future events by creating expectations of the prices of the assets that compose them. They anticipate the impact that new information entering into the market may have. 

No moving average dynamics were detected. In eleven cases, the model identification was MAR(1,1). In ten cases, MAR(0,2) and MAR(0,1) in three cases. No correlation exists between the residuals, as shown in Table \ref{tests}. In nineteen portfolios, we do not reject the null hypothesis of zero correlation for the test between the levels and squares of the errors. This means there are no remaining traces of GARCH effects in the residuals. In contrast, we rejected the null hypothesis of no correlation between the levels and the absolute value of the residuals on fifteen occasions. 

\vskip 0.15cm
\textbf{Size and Book-to-Market Portfolios}: For the small size and low B/M portfolio, we identify an MAR(1,1). However, we reject the null hypothesis for the absolute value and squares tests. The model found in the second portfolio, Medium Size 1 and B/M 2, is an MAR(1,1). For the residuals of this model, the null hypothesis in absolute value and squares is not rejected. The Small size and High B/M portfolio is an MAR(1,1). In this case, the null hypothesis is not rejected for the absolute value. However, lag number two is significant at 5\% but not at the 10\% significance level. We do not reject the null hypothesis for any lag in the squares. The Big Size and Low B/M portfolio is an MAR(1,1). We do not reject the null hypothesis in any test. The Medium Size 2 and B/M 2 portfolio is an MAR(0,2). In this case, we do not reject the null hypothesis for the first lag of the absolute value. However, we reject it after that. On the other hand, we do not reject the null hypothesis for the squares. The Big Size and High B/M portfolio is MAR(0,1). We reject the null hypothesis for the absolute value and do not reject the null hypothesis for the squares.\\

\textbf{Size and Operating Profitability Portfolios}: The Small Size and Low OP portfolio is an MAR(1,1). For both tests, the null hypothesis is not rejected. The Medium Size 1 and OP 2 portfolio is MAR(1,1). The null hypothesis is rejected for the absolute value but not rejected for the squares. The Small Size and High OP portfolio is an MAR(1,1). The null hypothesis is not rejected for the two tests, however, lags two and three of the absolute value test are marginally significant at 5\% but not at 10\%. The Big Size and Low OP portfolio is an MAR(0,2). In this case, the null hypothesis is not rejected for the squares test. However, the null hypothesis is rejected for lags three onwards in the absolute value test. The Medium Size 2 and OP 2 and Big Size and High OP portfolios are also MAR(0,2). However, we reject the null hypothesis in the absolute value test for lag two onwards. In the test of squares, we do not reject the null hypothesis. 
\\

\textbf{Size and Investment Portfolios}: The portfolios Small Size and Low INV and Medium Size 1 and INV 2 are MAR(1,1). We do not reject the null hypothesis at any lag for both tests. The Small Size and High INV portfolio is also an MAR(1,1), although we reject the null hypothesis in the absolute value test for lags two onwards. The Big Size and Low INV portfolio is MAR(0,2), and the null hypothesis is not rejected for both tests. On the other hand, the portfolios Medium Size 2 and INV 2 and Big Size and High INV are MAR(0,2). We do not reject the null hypothesis for the test of squares. However, we reject the null hypothesis for the absolute value test.
\\

\textbf{Size and Momentum Portfolios}: The Small Size and Low MOM portfolio is an MAR(0,1). On the other hand, the Medium Size 1 and MOM 2 and Small Size and High MOM portfolios are MAR(0,2). We reject the null hypothesis in the absolute value and squares tests for these three portfolios. The Big Size and Low MOM portfolio is an MAR(1,1), the Medium Size 2 and MOM 2 portfolio is an MAR(0,1), and the Big Size and high MOM portfolio is an MAR(0,2). For the residuals of the three portfolios, we reject the null hypothesis in the absolute value test for all lags. Conversely, we do not reject the null hypothesis in the test of squares.\\

\begin{landscape}
\begin{table}[h]
\centering
\tiny
\caption{Ljung-Box, $C_{\varepsilon,\lvert \varepsilon\rvert}$ and $C_{\varepsilon,\varepsilon^2}$ tests p-values}
\vskip 0.2cm
\label{tests}
\begin{tabular}{lccccccccccccccc}
\hline
\multicolumn{1}{c}{\textit{\textbf{Portfolios}}} & \multicolumn{5}{c}{Ljung-Box lags} & \multicolumn{5}{c}{$C_{\varepsilon,\lvert \varepsilon\rvert}$ lags} & \multicolumn{5}{c}{$C_{\varepsilon,\varepsilon^2}$ lags} \\ \hline
\textit{\textbf{Size and Book-to-Market}} & 1 & 2 & 3 & 4 & \multicolumn{1}{c|}{5} & 1 & 2 & 3 & 4 & \multicolumn{1}{c|}{5} & 1 & 2 & 3 & 4 & 5 \\ \hline
Small Size and Low B/M & 0.286 & 0.171 & 0.279 & 0.413 & \multicolumn{1}{c|}{0.537} & 0.000 & 0.000 & 0.000 & 0.000 & \multicolumn{1}{c|}{0.000} & 0.000 & 0.000 & 0.000 & 0.000 & 0.000 \\
Medium Size 1 and B/M 2 & 0.619 & 0.877 & 0.958 & 0.975 & \multicolumn{1}{c|}{0.991} & 0.482 & 0.400 & 0.564 & 0.663 & \multicolumn{1}{c|}{0.487} & 0.495 & 0.826 & 0.955 & 0.988 & 0.988 \\
Small Size and High B/M & 0.600 & 0.438 & 0.624 & 0.754 & \multicolumn{1}{c|}{0.827} & 0.534 & 0.027 & 0.050 & 0.097 & \multicolumn{1}{c|}{0.130} & 0.488 & 0.265 & 0.469 & 0.675 & 0.814 \\
Big Size and Low B/M & 0.343 & 0.518 & 0.711 & 0.691 & \multicolumn{1}{c|}{0.809} & 0.264 & 0.104 & 0.051 & 0.082 & \multicolumn{1}{c|}{0.086} & 0.184 & 0.229 & 0.317 & 0.443 & 0.629 \\
Medium Size 2 and B/M 2 & 0.628 & 0.288 & 0.407 & 0.392 & \multicolumn{1}{c|}{0.495} & 0.052 & 0.027 & 0.005 & 0.001 & \multicolumn{1}{c|}{0.002} & 0.263 & 0.201 & 0.152 & 0.161 & 0.281 \\
Big Size and High B/M & 0.193 & 0.366 & 0.513 & 0.596 & \multicolumn{1}{c|}{0.709} & 0.047 & 0.003 & 0.003 & 0.003 & \multicolumn{1}{c|}{0.006} & 0.098 & 0.157 & 0.279 & 0.372 & 0.546 \\ \hline
 & \multicolumn{5}{c}{Ljung-Box lags} & \multicolumn{5}{c}{$C_{\varepsilon,\lvert \varepsilon\rvert}$ lags} & \multicolumn{5}{c}{$C_{\varepsilon,\varepsilon^2}$ lags} \\ \hline
\textit{\textbf{Size and Operating Profitability}} & 1 & 2 & 3 & 4 & \multicolumn{1}{c|}{5} & 1 & 2 & 3 & 4 & \multicolumn{1}{c|}{5} & 1 & 2 & 3 & 4 & 5 \\ \hline
Small  Size and Low OP & 0.232 & 0.420 & 0.561 & 0.703 & \multicolumn{1}{c|}{0.769} & 0.197 & 0.218 & 0.370 & 0.497 & \multicolumn{1}{c|}{0.395} & 0.184 & 0.432 & 0.659 & 0.828 & 0.902 \\
Medium Size 1 and OP 2 & 0.180 & 0.178 & 0.304 & 0.439 & \multicolumn{1}{c|}{0.565} & 0.101 & 0.018 & 0.038 & 0.064 & \multicolumn{1}{c|}{0.029} & 0.085 & 0.099 & 0.217 & 0.388 & 0.519 \\
Small Size and High OP & 0.456 & 0.623 & 0.804 & 0.872 & \multicolumn{1}{c|}{0.860} & 0.057 & 0.036 & 0.046 & 0.074 & \multicolumn{1}{c|}{0.011} & 0.086 & 0.218 & 0.379 & 0.573 & 0.526 \\
Big Size and Low OP & 0.860 & 0.551 & 0.749 & 0.778 & \multicolumn{1}{c|}{0.690} & 0.286 & 0.224 & 0.064 & 0.013 & \multicolumn{1}{c|}{0.019} & 0.341 & 0.377 & 0.393 & 0.403 & 0.468 \\
Medium Size 2 and OP 2 & 0.669 & 0.355 & 0.558 & 0.539 & \multicolumn{1}{c|}{0.636} & 0.063 & 0.028 & 0.018 & 0.002 & \multicolumn{1}{c|}{0.003} & 0.274 & 0.277 & 0.346 & 0.318 & 0.477 \\
Big Size and High OP & 0.500 & 0.262 & 0.355 & 0.362 & \multicolumn{1}{c|}{0.409} & 0.261 & 0.042 & 0.004 & 0.001 & \multicolumn{1}{c|}{0.002} & 0.427 & 0.119 & 0.073 & 0.079 & 0.139 \\ \hline
 & \multicolumn{5}{c}{Ljung-Box lags} & \multicolumn{5}{c}{$C_{\varepsilon,\lvert \varepsilon\rvert}$ lags} & \multicolumn{5}{c}{$C_{\varepsilon,\varepsilon^2}$ lags} \\ \hline
\textit{\textbf{Size and Investment}} & 1 & 2 & 3 & 4 & \multicolumn{1}{c|}{5} & 1 & 2 & 3 & 4 & \multicolumn{1}{c|}{5} & 1 & 2 & 3 & 4 & 5 \\ \hline
Small   Size and Low INV & 0.286 & 0.305 & 0.498 & 0.665 & \multicolumn{1}{c|}{0.775} & 0.128 & 0.083 & 0.127 & 0.221 & \multicolumn{1}{c|}{0.134} & 0.259 & 0.370 & 0.545 & 0.750 & 0.867 \\
Medium Size 1 and INV 2 & 0.321 & 0.579 & 0.778 & 0.851 & \multicolumn{1}{c|}{0.928} & 0.184 & 0.187 & 0.391 & 0.415 & \multicolumn{1}{c|}{0.271} & 0.128 & 0.376 & 0.642 & 0.793 & 0.893 \\
Small Size and High INV & 0.266 & 0.342 & 0.504 & 0.654 & \multicolumn{1}{c|}{0.705} & 0.128 & 0.027 & 0.036 & 0.077 & \multicolumn{1}{c|}{0.035} & 0.058 & 0.085 & 0.203 & 0.375 & 0.419 \\
Big Size and Low INV & 0.460 & 0.294 & 0.334 & 0.384 & \multicolumn{1}{c|}{0.371} & 0.265 & 0.083 & 0.142 & 0.056 & \multicolumn{1}{c|}{0.088} & 0.398 & 0.309 & 0.376 & 0.463 & 0.533 \\
Medium Size 2 and INV 2 & 0.672 & 0.224 & 0.381 & 0.499 & \multicolumn{1}{c|}{0.459} & 0.120 & 0.036 & 0.006 & 0.001 & \multicolumn{1}{c|}{0.002} & 0.267 & 0.178 & 0.169 & 0.232 & 0.300 \\
Big Size and High INV & 0.410 & 0.346 & 0.548 & 0.606 & \multicolumn{1}{c|}{0.743} & 0.009 & 0.003 & 0.000 & 0.000 & \multicolumn{1}{c|}{0.001} & 0.084 & 0.059 & 0.083 & 0.088 & 0.181 \\ \hline
 & \multicolumn{5}{c}{Ljung-Box lags} & \multicolumn{5}{c}{$C_{\varepsilon,\lvert \varepsilon\rvert}$ lags} & \multicolumn{5}{c}{$C_{\varepsilon,\varepsilon^2}$ lags} \\ \hline
\textit{\textbf{Size and Momentum}} & 1 & 2 & 3 & 4 & \multicolumn{1}{c|}{5} & 1 & 2 & 3 & 4 & \multicolumn{1}{c|}{5} & 1 & 2 & 3 & 4 & 5 \\ \hline
Small Size and Low MOM & 0.381 & 0.587 & 0.784 & 0.876 & \multicolumn{1}{c|}{0.928} & 0.001 & 0.000 & 0.000 & 0.000 & \multicolumn{1}{c|}{0.000} & 0.000 & 0.001 & 0.003 & 0.007 & 0.007 \\
Medium Size 1 and MOM 2 & 0.505 & 0.788 & 0.924 & 0.933 & \multicolumn{1}{c|}{0.971} & 0.002 & 0.001 & 0.002 & 0.003 & \multicolumn{1}{c|}{0.003} & 0.009 & 0.027 & 0.082 & 0.158 & 0.231 \\
Small Size and High MOM & 0.398 & 0.390 & 0.597 & 0.756 & \multicolumn{1}{c|}{0.864} & 0.001 & 0.000 & 0.000 & 0.000 & \multicolumn{1}{c|}{0.000} & 0.002 & 0.004 & 0.012 & 0.023 & 0.032 \\
Big Size and Low MOM & 0.835 & 0.183 & 0.326 & 0.482 & \multicolumn{1}{c|}{0.625} & 0.161 & 0.003 & 0.003 & 0.007 & \multicolumn{1}{c|}{0.014} & 0.122 & 0.054 & 0.052 & 0.112 & 0.207 \\
Medium Size 2 and MOM 2 & 0.243 & 0.351 & 0.532 & 0.626 & \multicolumn{1}{c|}{0.760} & 0.010 & 0.001 & 0.001 & 0.001 & \multicolumn{1}{c|}{0.001} & 0.109 & 0.067 & 0.077 & 0.138 & 0.263 \\
Big Size and high MOM & 0.464 & 0.764 & 0.891 & 0.589 & \multicolumn{1}{c|}{0.530} & 0.337 & 0.078 & 0.031 & 0.003 & \multicolumn{1}{c|}{0.003} & 0.239 & 0.414 & 0.493 & 0.341 & 0.410 \\ \hline
\end{tabular}
\end{table}
\end{landscape}
\section{Conclusions}

In this study, we introduced a novel estimation method for MARMA models that may contain both causal and non-causal components, as well as invertible and non-invertible elements. Our approach relies on the spectrum, bispectrum, periodogram, and biperiodogram. Instead of leveraging the i.i.d. nature of residuals as many methods do, we utilize high-order cumulants. The proposed estimation function yields unbiased parameters and displays an asymptotic normal distribution. It consistently identifies the underlying data-generating process, particularly when the data shows deviations from normality. For this identification, it's crucial for the third and fourth moments/cumulants to be non-zero. Our claims are backed by comprehensive Monte Carlo simulations using the alpha-stable and skew-t distribution.

Additionally, we introduce a simulation technique for MARMA processes that utilizes the Fourier transform of the error sequence and the transfer function. This method stands out because it avoids the AR or MA representation of the stationary MARMA solution. As a result, no data is lost due to the need for an initial or final condition, and it eliminates potential bias from using lead variables and future errors.

When conducting estimations in the frequency domain, errors can still be extracted in the time domain using the inverse Fourier transform. We suggest three tests on these residuals to assess their potential dependence structures and gauge how closely they align with the i.i.d. assumption. 

Lastly, we applied our methods to an empirical dataset comprising 24 monthly returns of Famma-French stock portfolios from emerging markets. We identified a mix of causal, non-causal, and purely non-causal dynamics. This result indicates that these portfolios have the characteristic of adjusting adequately to near future events by creating expectations of the prices of the assets that compose them. They anticipate the impact that new information entering into the market may have. This feature is especially valuable from the point of view of risk management and portfolio construction. 

The calculated residuals consistently resembled white noise without any heteroscedastic effects. In contrast, estimating using the conventional AR process, which assumes causality, did produce white noise residuals. However, these also showed heteroscedastic effects, suggesting a need for a GARCH model.

\bibliographystyle{johd}
\bibliography{bib}

\subsection*{Appendix A: Asymptotic Variance}\label{asymnorm}

Holding Theorem 1 and 3 in \cite{velasco2018frequency}, the asymptotic normality of the second and third-order estimator can be expressed as 

    \[\sqrt{T}\left(\boldsymbol{\hat{\vartheta}}-\vartheta_0\right)\rightarrow N(0,\mathbb{W}).
\]

The matrix $\mathbb{W}$ stands for the asymptotic variance

\[\mathbb{W}=\begin{bmatrix}\left(\Phi_0+\Phi_0^{*}\right)^{-1} & \Phi_0^{-1} \\\Phi_0^{-1} & \frac{v_4+2}{v_3^2}\Phi_0^{-1}+\Phi_0^{-1}\Phi_0^{*} \Phi_0^{-1}\end{bmatrix},
\]

where, $v_4=\frac{\kappa_4}{\kappa_2^{2}}$ stands for the standardized kurtosis, 
$v_3=\frac{\kappa_3}{\kappa_2^{3/2}}$ stands for the standardized skewness, and
\[\Phi_0=\frac{1}{2\pi}\int_{-\pi}^{\pi}\psi^1(\vartheta_0;\omega)\psi^1(\vartheta_0;-\omega)'d\omega;\;\;\Phi_0^{*}=\frac{1}{2\pi}\int_{-\pi}^{\pi}\psi^1(\vartheta_0;\omega)\psi^1(\vartheta_0;\omega)' d\omega, \]

where $\psi^1(\vartheta_0;\omega)=\psi(\vartheta_0;\omega)-\mu(\vartheta_0)$,  $\psi(\vartheta_0;\omega)=\frac{\partial}{\partial \vartheta}log\;\psi(\vartheta_0;\omega)$, and $\mu(\vartheta_0)=\frac{1}{2\pi}\int_{-\pi}^{\pi}\psi(\vartheta_0;\omega)d\omega$.To prove the existence of $\mu(\vartheta_0)$, one needs to exclude the unit root case. This integral is different from $0$ only when the model is noninvertible. The matrix $\Phi_0$ is positive definite, excluding unit and common roots in the MA and AR polynomials. For the second order estimation the asymptotic variance is $\mathbb{W}_{1,1}=\left(\Phi_0+\Phi_0^{*}\right)^{-1}$. For the third order estimation is $\mathbb{W}_{2,2}=\frac{v_4+2}{v_3^2}\Phi_0^{-1}+\Phi_0^{-1}\Phi_0^{*} \Phi_0^{-1}$.  

For an MA(1,0), the transfer function is
$\psi(\vartheta_0;\omega)=1+\theta^{+}_1 e^{i\omega}$. The FOC respect with the parameter is
\[\varphi(\vartheta_0;\omega)=\frac{\partial}{\partial \vartheta}log \left(1+\theta^{+}_1 e^{i\omega}\right)=\frac{e^{i\omega}}{1+\theta^{+}_1 e^{i\omega}},\]
and 
\[\varphi(\vartheta_0;-\omega)=\frac{e^{-i\omega}}{1+\theta^{+}_1 e^{-i\omega}}.\]

The centering parameters can be found by the Cauchy formula as
\[\mu(\vartheta_0, \omega)=\frac{1}{2\pi}\int_{-\pi}^{\pi}\frac{e^{i\omega}}{1+\theta^{+}_1 e^{i\omega} }d\omega=\begin{cases}\frac{1}{\theta^{+}_1} & \lvert\theta^{+}_1\rvert > 1\\0 & \end{cases} \; if\; \frac{1}{\lvert\theta^{+}_1\rvert}\neq1,\]

\[\mu(\vartheta_0,-\omega)=\frac{1}{2\pi}\int_{-\pi}^{\pi}\frac{e^{-i\omega}}{1+\theta^{+}_1 e^{-i\omega} }d\omega=\begin{cases}\frac{1}{\theta^{+}_1} & \lvert\theta^{+}_1\rvert \geq 1\\0 & \end{cases} \; if\; \frac{1}{\lvert\theta^{+}_1\rvert}\neq1.\]

Then

\[\varphi^1(\vartheta_0,\omega)=\frac{e^{i\omega}}{1+\theta^{+}_1 e^{i\omega}};\;\;\varphi^1(\vartheta_0,-\omega)=\frac{e^{-i\omega}}{1+\theta^{+}_1 e^{-i\omega}}.\]

In this sense,

\[\Phi_0=\frac{1}{2\pi}\int_{-\pi}^{\pi} \frac{e^{i\omega}}{1+\theta^{+}_1 e^{i\omega}}\frac{e^{-i\omega}}{1+\theta^{+}_1 e^{-i\omega}}d\omega=\frac{1}{1-\theta^{+2}_1}, \;\; \textrm{and}\;\;\Phi^{*}_0=\frac{1}{2\pi}\int_{-\pi}^{\pi} \frac{e^{i\omega}}{1+\theta^{+}_1 e^{i\omega}}\frac{e^{i\omega}}{1+\theta^{+}_1 e^{i\omega}}d\omega=0.\]

For other MARMA models, the procedure is straightforward. \cite{lobato2022single} present closed forms for the standard errors, greatly simplifying the inference.

\section*{Figures}

\end{document}